\documentclass[journal, twocolumn]{IEEEtran}

\usepackage[usenames,dvipsnames]{xcolor}
\usepackage{amsmath,amssymb,amsfonts,tikz,pgfplots,bm}
\usepackage{cite,booktabs,tabularx}
\usepackage{algorithmic}
\usepackage{graphicx}
\usepackage{textcomp}
\usepackage{verbatim}
\usepackage{amsmath}
\usepackage[ruled,vlined]{algorithm2e}
\usepackage{nicefrac}
\usepackage[inline]{enumitem}
\usepackage{array}
\usepackage{balance}

\newcommand{\len}{\ensuremath{N}}
\newcommand{\leni}{\ensuremath{n}}
\newcommand{\mem}{\ensuremath{m}}
\newcommand{\dimi}{\ensuremath{k}}
\newcommand{\dimibit}{\ensuremath{k}}
\newcommand{\leno}{\ensuremath{N_\mathsf{o}}}
\newcommand{\dimo}{\ensuremath{K}}
\newcommand{\numS}{\ensuremath{M}}
\newcommand{\bl}{\leno+m}
\newcommand{\rateo}{\ensuremath{R_\mathsf{o}}}
\newcommand{\ratei}{\ensuremath{R_\mathsf{i}}}

\newcommand{\D}{\mathsf{D}}
\newcommand{\I}{\mathsf{I}}
\newcommand{\T}{\mathsf{T}}
\renewcommand{\S}{\mathsf{S}}

\let\vv\v

\newcommand{\e}{\ensuremath{\mathrm{e}}}
\renewcommand{\d}{\ensuremath{\boldsymbol{d}}}
\newcommand{\s}{\ensuremath{\boldsymbol{s}}}
\renewcommand{\u}{\ensuremath{\boldsymbol{u}}}
\newcommand{\w}{\ensuremath{\boldsymbol{w}}}
\newcommand{\x}{\ensuremath{\boldsymbol{x}}}
\newcommand{\y}{\ensuremath{\boldsymbol{y}}}
\renewcommand{\v}{\ensuremath{\boldsymbol{v}}}

\newcommand{\Al}{\ensuremath{\Sigma}}
\newcommand{\F}{\ensuremath{\mathbb{F}}}

\usetikzlibrary{positioning,calc}
\tikzstyle{block} = [rectangle, draw, text centered, rounded corners, minimum height=1.5em,fill=black!5!]

\def\approxprop{%
	\def\p{%
		\setbox0=\vbox{\hbox{$\propto$}}%
		\ht0=0.6ex \box0 }%
	\def\s{%
		\vbox{\hbox{$\sim$}}%
	}%
	\mathrel{\raisebox{0.7ex}{%
			\mbox{$\underset{\s}{\p}$}%
	}}%
}

\usepackage{soul}
\newcommand{\al}[1]{{\footnotesize  [\hl{#1} \textcolor{blue!60!black}{--Andreas}]\normalsize}}

\newcommand{\field}[1]{\ensuremath{\mathbb{F}_{#1}}}
\newcommand{\vecspace}[2]{\ensuremath{\mathbb{F}_{#1}^{#2}}}
\newcommand{\outq}{\ensuremath{q_\mathsf{o}}}
\newcommand{\kernel}{\ensuremath{\boldsymbol{K}}}
\newcommand{\oinffrz}{\ensuremath{\boldsymbol{u}^{\prime}}}
\newcommand{\oinfnofrz}{\ensuremath{\boldsymbol{u}}}
\newcommand{\olen}{\ensuremath{N_\mathsf{o}}}
\newcommand{\odim}{\ensuremath{K}}
\newcommand{\ocw}{\ensuremath{\boldsymbol{w}}}
\newcommand{\crclen}{\ensuremath{\ell_\mathrm{CRC}}}
\newcommand{\chout}{\ensuremath{\boldsymbol{y}}}
\newcommand{\numseq}{\ensuremath{M}}

\newcommand{\tv}{\ensuremath{t}}

\begin{document}

\title{Concatenated Codes for Multiple Reads\\ of a DNA Sequence}

\author{\IEEEauthorblockN{{\bf Issam Maarouf}\IEEEauthorrefmark{2}, {\bf Andreas Lenz}\IEEEauthorrefmark{1}, {\bf Lorenz Welter}\IEEEauthorrefmark{1},\\ {\bf Antonia Wachter-Zeh}\IEEEauthorrefmark{1}, {\bf Eirik Rosnes}\IEEEauthorrefmark{2}, {\bf and Alexandre Graell i Amat}\IEEEauthorrefmark{3}\IEEEauthorrefmark{2}}

\IEEEauthorblockA{
	\IEEEauthorrefmark{2}Simula UiB, N-5006  Bergen,  Norway
}

\IEEEauthorblockA{
	\IEEEauthorrefmark{1}Institute for Communications Engineering, Technical University of Munich, DE-80333 Munich, Germany
}

\IEEEauthorblockA{
	\IEEEauthorrefmark{3}Department of Electrical Engineering, Chalmers University of Technology, SE-41296 Gothenburg, Sweden
}

\thanks{
Parts of this work have been presented at the 2020/2021 IEEE Information Theory Workshop (ITW) \cite{LenzMaaroufConcatenatedCodes_DNA2020}.

The work of A. Lenz, L. Welter, and A. Wachter-Zeh has been supported by the European Research
Council (ERC) under the European Union’s Horizon 2020 research and innovation programme (Grant Agreement No. 801434).

The work of A. Graell i Amat was supported by the Swedish Research Council under grant 2020-03687 and by a Technical University of Munich Global Visiting Professor Fellowship.}}

\maketitle

\begin{abstract}
Decoding sequences that stem from multiple transmissions of a codeword over an insertion, deletion, and substitution channel is a critical component of efficient deoxyribonucleic acid (DNA) data storage systems. In this paper, we consider a concatenated coding scheme with an outer nonbinary low-density parity-check code  or a polar code and either an inner convolutional code or a time-varying block code. We propose two novel decoding algorithms for inference from multiple received sequences, both combining the inner code and channel to a joint hidden Markov model to infer symbolwise a posteriori probabilities (APPs). The first decoder computes the exact APPs by jointly decoding the received sequences, whereas the second decoder approximates the APPs by combining the results of separately decoded received sequences and has a complexity that is linear with the number of sequences. Using the proposed algorithms, we evaluate the performance of decoding multiple received sequences by means of achievable information rates and Monte-Carlo simulations. We show significant performance gains compared to a single received sequence. In addition, we succeed in improving the performance of the aforementioned coding scheme by optimizing both the inner and outer codes.
\end{abstract}


\IEEEpeerreviewmaketitle


\section{Introduction} \label{sec:introduction}
Error correction of data storage in deoxyribonucleic acid (DNA) has recently gained a lot of attention from the coding theory community. This attention increased after several successful experiments \cite{church_next-generation_2012,goldman_towards_2013,grass_robust_2015,blawat_forward_2016,bornholt_dna-based_2016,erlich_dna_2017,yazdi_portable_2017,organick_random_2018,wang_high_2019,chandak_overcoming_2020,yazdi_rewritable_2015,pan_rewritable_2021,tabatabaei_expanding_2021,antkowiak_low_2020} that demonstrated the viability of using synthetic DNA as a reliable medium for data storage. As a result of the pioneering DNA storage experiments, several information-theoretic problems have been identified. The most important problem to our work is  reliable communication over channels that introduce insertions, deletions, and substitutions (IDSs)~\cite{heckel_characterization_2019} as the processes of DNA synthesis and DNA sequencing introduce errors in the forms of IDSs. Furthermore, in the literature, channels that introduce IDSs have been proposed to model synchronization errors. Thus, coding techniques are an indispensable component to cope with IDSs and improve the reliability of DNA storage systems and channels that are prone to desynchronization.

Work on synchronization errors began decades ago. Several papers in the 1960s-70s have dealt with information-theoretic aspects of IDS errors, some even proposed codes to correct these errors \cite{gallager_sequential_1961,zigangirov_sequential_1969,calabi_general_1969, bahl_decoding_1975, levenshtein_binary_1966}. From these works, several constructions of error-correcting codes for the IDS channel have been proposed in the last decade. Among the most important ones, and most relevant to our work, is the one introduced by Davey and MacKay \cite{davey_reliable_2001}. In that paper, the authors introduce a concatenated coding scheme composed of an inner block code and an outer low-density parity-check (LDPC) code. In addition, they propose a decoding algorithm based on dynamic programming that represents the inner code and channel by a hidden Markov model (HMM). By doing so, the decoding algorithm allows to infer a posteriori probabilities (APPs) to be passed to the outer decoder, which will complete the message recovery process. Inspired by Davey and MacKay's work, the Markov process of convolutional codes  was extended to the IDS channel, allowing for decoding algorithms of convolutional codes to be run for the IDS channel \cite{mansour_convolutional_2010,buttigieg_improved_2015}. An improvement of the decoding algorithm in \cite{davey_reliable_2001} was introduced in \cite{briffa_improved_2010}. Furthermore, marker codes were used as inner codes in \cite{inoue_adaptive_2012},  which improved the performance of the inner codes of \cite{davey_reliable_2001}. Additionally, standalone codes (i.e., without  an inner code) such as LDPC codes in \cite{shibata_design_2019} and polar codes in \cite{Koremura2020, Pfister2021} were studied to tackle synchronization errors.

Most of these studies have focused on error correction for a single block transmission over an IDS channel. However, in DNA-based storage the data is  synthesized into many short DNA strands (or sequences), where each  strand is replicated thousands of times. As a result, when the data is read via DNA sequencing, multiple copies of each data strand are obtained. 

Decoding multiple reads of a DNA sequence is typically performed via a multiple sequence alignment (MSA) \cite{edgar_muscle_2004,kim_psar-align_2014,notredame2000tcoffee,Reinert2017seqan} of the received sequences, followed by a majority decision on the alignment. This method is popular, since it is possible to employ standard decoders on the single resulting sequence to correct remaining errors and has been used in several recent experiments \cite{yazdi_portable_2017,organick_random_2018,pan_rewritable_2021,antkowiak_low_2020}.

In this paper, we introduce concatenated coding schemes for multiple sequence transmission over parallel IDS channels over the DNA alphabet. The proposed schemes employ an  inner block code or convolutional code (as introduced in \cite{davey_reliable_2001} and \cite{mercier_convolutional_2009}, respectively) that corrects insertions and deletions, concatenated with an outer LDPC or polar code, which corrects remaining (mostly substitution) errors. Our key contribution is a low-complexity decoding strategy that properly combines information from the multiple reads. Compared to MSA techniques, which make hard decisions on the DNA symbols, the proposed strategy provides soft information to the outer decoder. The proposed schemes achieve significant gains over the single sequence transmission case and outperform MSA if the number of traces is small. Furthermore, for a larger number of traces, they yield comparable performance to MSA at lower complexity. Our main contributions are summarized as follows.

\begin{itemize}
  \item We propose two novel decoding algorithms for the decoding of the combination of the inner code and the IDS channel with multiple reads:  an exact symbolwise maximum a posteriori (MAP) decoder based on a multidimensional trellis encompassing  the inner code and the multiple  IDS channels, and a sub-optimal decoder that decodes each received sequence independently and combines the results into a single sequence of approximate APPs that are fed to the outer decoder. The first decoder is optimal but of high complexity, whereas the second decoder is a  practical decoder that significantly reduces the complexity while achieving excellent performance. For instance,  for a considered scenario, we show that separate decoding with three sequences has comparable performance  to joint decoding with two sequences, but with much lower decoding complexity.
  \item We improve on the performance of  earlier concatenated  schemes by providing optimization techniques tailored to the IDS channel for both the inner and outer codes. In particular, we design an inner time-varying code (TVC) concatenated with either an LDPC code or a polar code. 
   \item To gain  insight into the asymptotic performance of the proposed schemes, we compute achievable information rates (AIRs) for the case with no iterations between the inner and outer decoder and the case where iterations are allowed. The computed AIRs depend on the inner code\textemdash but not on the outer code\textemdash and thus can be used to evaluate and select the inner code.
\end{itemize}

\subsection*{Related Work}

Work related to ours includes the study of repeated transmission over an erroneous channel, which goes back to Levenshtein \cite{levenshtein_efficient_2001}, who introduced the sequence reconstruction problem. Here, the goal is to analytically quantify the number of sequences, as a function of the sequence length, that are required to guarantee correct reconstruction under an adversarial channel. Recently, the original formulation has been extended to an adversarial DNA channel \cite{sini_reconstruction_2019}. Furthermore, similar in spirit to our work, yet different in methods and objectives, is the research on the trace reconstruction problem over deletion channels \cite{brakensiek_coded_2020,cheraghchi_coded_2020,abroshan_coding_2019}, which is the probabilistic variant of the sequence reconstruction problem. These works focus on asymptotic results, while we discuss both asymptotic and finite-length results of received sequences over a channel that additionally allows insertions and substitutions. More recently, the trace reconstruction problem has also been formulated for a fixed number of sequences with a larger focus on algorithmic aspects \cite{srinivasavaradhan_symbolwise_2019,sabary_reconstruction_2020}. However, these works consider only uncoded  sequences, while we are interested in coded transmissions. 

Inspired by  our work \cite{LenzMaaroufConcatenatedCodes_DNA2020}, independently of this paper, 
the authors in \cite{Srinivasavaradhan2021TrellisBMA}  worked on multiple sequence transmission over parallel IDS channels. In particular in \cite{Srinivasavaradhan2021TrellisBMA}, they presented a practical sub-optimal decoder, referred to as trellis bitwise majority alignment,  which slightly improves the performance compared to our sub-optimal decoder, while slightly increasing the complexity.  The authors in \cite{Srinivasavaradhan2021TrellisBMA} also proposed a less complex decoder than our optimal APP decoder. In \cite{SakogawaMAP_IDS_2020}, a more complex sub-optimal decoder compared to our separate decoder was presented for multiple sequence transmission over the IDS channel, as well as a similar optimal APP decoder as ours. In \cite{Shibata2020}, a concatenated coding scheme with an outer LDPC code and an inner trellis code approaching the Markov capacity of a channel with insertions and deletions for single sequence transmission was proposed.

Other works that discuss AIRs for IDS channels  are   \cite{Fertonani2011,Shibata2020}. These works consider a binary channel and the single sequence case. In \cite{Fertonani2011}, bounds on the capacity of binary IDS channels are computed, while in \cite{Shibata2020} a Markovian input process for their binary insertion and deletion channel is optimized, which results in better AIRs. 


\section{System Model} \label{sec:system}

\subsection{Channel Model}

We consider the  IDS channel model depicted in Fig.~\ref{fig:ids:channel} \cite{Dobrushin1967,davey_reliable_2001, briffa_improved_2010, mansour_convolutional_2010, inoue_adaptive_2012,buttigieg_improved_2015, Koremura2020, Pfister2021}. Let $\x = (x_1,\ldots ,x_\len)$, $x_i \in \Al _q = \{0,1,\dots,q-1\}$, be the information sequence to be transmitted over the channel. The sequence can be viewed as a queue where each symbol $x_i$ is successively transmitted over the channel. We describe in the following how the received sequence $\y = (y_1,\ldots,y_{\len'})$ is generated state by state. Assume $x_i$ is queued and therefore sought to be transmitted over the channel. The channel enters state $x_i$ and three events may occur: \begin{enumerate*}[label=(\roman*)] \item With probability $p_\I$, an insertion event occurs where a uniformly random symbol $a \in \Al_q$ is appended to the received sequence. In this case, $x_i$ remains in the queue and the channel returns to state $x_i$. \item Symbol $x_i$ is deleted with probability $p_\D$. Consequently, the queued symbol $x_i$ is not appended to $\y$ and the next symbol $x_{i+1}$ is enqueued and the channel enters state $x_{i+1}$. \item $x_i$ is transmitted with probability $p_\T = 1-p_\I -p_\D$. In this case, the symbol is substituted with a uniformly random symbol $a^* \neq x_i$ with probability $p_\S$ and the next transmit symbol $x_{i+1}$ is enqueued and the channel enters state $x_{i+1}$. \end{enumerate*} When the last transmit symbol $x_{\len}$ leaves the queue, the procedure finishes and the channel outputs $\y$. Note that the length of the output sequence $\len'$ is random and depends on the probabilities $p_\D$ and $p_\I$.

\subsection{Coding Scheme}
Following \cite{davey_reliable_2001}, we consider an error-correcting code consisting of the serial concatenation of two codes. The  role of the inner code is to maintain synchronization and provide reliable soft information to the outer code, while the task of the outer code is to use this soft information to perform an accurate estimate of the transmitted information. In particular, given that the inner code provides reliable synchronization information, the outer code corrects the remaining errors on the synchronized sequence.

We investigate multiple choices of inner codes, which can be classified into two categories: (i) Convolutional codes as introduced in \cite{mercier_convolutional_2009}, and (ii) TVCs. Note that the inner code construction introduced in \cite{davey_reliable_2001} can be seen as a non-time-varying block code or a convolutional code with memory zero and thus, for simplicity, we restrict some passages in this work to convolutional inner codes only. Moreover, the outer code is either a nonbinary LDPC  code or a nonbinary polar code. 

The considered system model is depicted in Fig.~\ref{fig:system:model}. The information vector $\u = (u_1, \ldots , u_\dimo)$,  $u_i \in \field{\outq}$, is encoded by an $[\leno,\dimo]_{\outq}$ outer code to a codeword $\ocw = (w_1,\dots,w_{\leno})$, $w_i \in \field{{\outq}}$. We consider the field $\field{\outq}$ to be a binary extension field of size $\outq = 2^\dimi$, where $\dimi$ is the dimension of the inner code.  The codeword $\w$ is then encoded by the inner code, either  an $(\leni,\dimi,\mem)_q$ convolutional code of block length $\leni$, binary input length $\dimibit$, memory $\mem$, and output alphabet $\Sigma_q$, or an $[\leni, \dimi,\tv]_q$ TVC, where $\tv$ denotes the number of different codebooks that are used.  The  codeword generated by the inner code is denoted by $\v = (v_1,\ldots,v_\len)$, $v_i \in \Sigma_q$, of length  $\len = (\olen+m)\leni$ for the case of an inner convolutional code (due to termination) and  $\len = \olen \leni$ for the case of an inner TVC. Finally, a pseudo-random offset sequence is optionally added to $\v$ before transmission, resulting in the sequence $\x = (x_1,\ldots,x_\len)$. The random sequence is known to the decoder and supports the inner decoder to maintain synchronization at especially high probabilities of insertions or deletions \cite{davey_reliable_2001,Buttigieg_CodebookAM_2011}. Section \ref{sec:inner:code} gives a more detailed description of the role of the random sequence. The rate of the concatenated code is measured in bits per DNA symbol and is given as $R = \rateo \ratei = \nicefrac{\dimo \dimibit}{\len}$, where  $\rateo = \nicefrac{\dimo}{\leno}$ is the rate of the outer code  and $\ratei = \nicefrac{\olen \dimibit}{\len}$ the rate of the inner code.

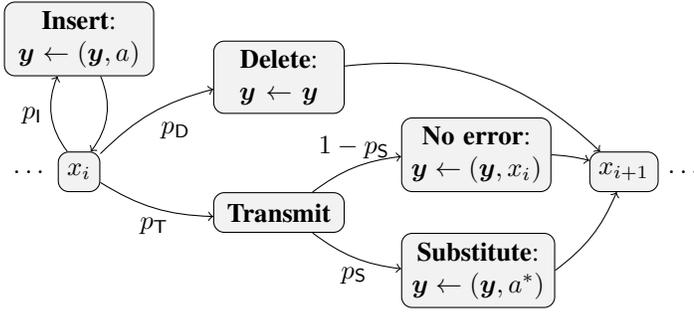
\begin{figure}[t]
	\centering
	\begin{tikzpicture}
		\node[block] (xi) {$x_i$};
		\node[left=0cm of xi] {$\dots$};
		
		\node[block, above=1cm of xi, text width=1.75cm] (ins) {{\bf Insert}: $\y \leftarrow (\y,a)$};
		
		\node[block, above right=.5cm and 1.5cm of xi, text width=1.5cm] (del) {{\bf Delete}: $\y \leftarrow \y$};
		
		\node[block, below right=.0cm and 1.5cm of xi, text width=1.5cm] (tr) {{\bf Transmit}};
		
		\node[block, above right=.0cm and .75cm of tr, text width=1.75cm] (noerr) {{\bf No error}: $\y \leftarrow (\y,x_i)$};
		\node[block, below right=.0cm and .75cm of tr, text width=1.8cm] (sub) {{\bf Substitute}: $\y \leftarrow (\y,a^*)$};
		
		\node[block, right=6.5cm of xi] (xip1) {$x_{i+1}$};
		\node[right=0cm of xip1] {$\dots$};
		
		\draw[->] (xi) to [bend left] node[left] {$p_\I$} (ins);
		\draw[->] (ins) to [bend left] (xi);
		
		\draw[->] (xi) to [bend left=15] node[below right] {$p_\D$} (del);
		
		\draw[->] (xi) to [bend right=15] node[below] {$p_\T$} (tr);
		
		\draw[->] (tr) to [bend right=15] node[below] {$p_\S$} (sub);
		\draw[->] (tr) to [bend left=15] node[above] {$1-p_\S$} (noerr);
		
		\draw[->] (del) to [bend left=25] (xip1);
		
		\draw[->] (noerr.east) to [bend left=5] (xip1);
		\draw[->] (sub.east) to [bend right=15] (xip1);
	\end{tikzpicture}
	
	\caption{State-based IDS channel \cite{davey_reliable_2001}. }
	\label{fig:ids:channel}
\end{figure}
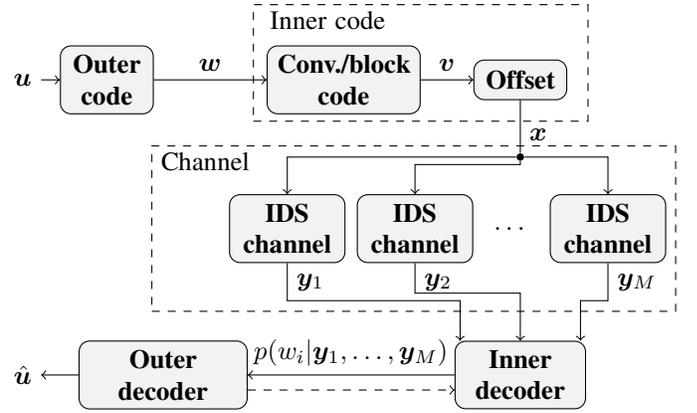
\begin{figure}[t]
	\centering
	\begin{tikzpicture}

		\node (u) {$\u$};
		
		\node[block, right=0.25cm of u,text width=1.cm] (out) {\bf Outer code};
		
		\node[block, right=1.5cm of out,text width=1.8cm] (inn) {\bf Conv./block code};
		
		\node[block, right=.7cm of inn,text width=1.cm] (off) {\bf Offset};
		
		\node[block, below left=1.25cm and 1.7cm of off, text width=1.3cm] (c1) {\bf IDS channel};
		\node[block, right = .15cm of c1, text width=1.3cm] (c2) {\bf IDS channel};
		\node[right = .15cm of c2] (cd) {$\dots$};
		\node[block, right = .15cm of cd, text width=1.3cm] (cA) {\bf IDS channel};
		
		\node[circle, fill, below=.7cm of off,inner sep=.035cm] (circ) {};
		
		\node[block,below=3.2cm of off, text width=1.5cm] (inndec) {\bf Inner decoder};
		
		\node[block,left=2.75cm of inndec, text width=2cm] (outdec) {\bf Outer decoder};
		
		\node (uh) at (u |- outdec) {$\hat{\u}$};
		
		\draw[->] (u) -- (out);
		\draw[->] (out) -- node[above] {$\w$} (inn);	
		
		\draw[->] (inn) -- node[above] {$\v$} (off);	
		
		\draw (off) -- node[right,pos=.65] {$\x$} (circ);
		\draw[->] (circ) -| (c1.north);
		\draw[->] (circ) -- +(0,-0.1) -| (c2.north);
		\draw[->] (circ) -| (cA.north);
		
		\draw[->] (c1.south) -- node[right] {$\y_1$} +(0,-0.5) -| ($(inndec.north) + (-0.8,0)$);
		\draw[->] (c2.south) -- node[right,pos=0.625] {$\y_2$} +(0,-0.4) -| (inndec.north);
		\draw[->] (cA.south) -- node[right] {$\y_{\numS}$} +(0,-0.5) -| ($(inndec.north) + (0.8,0)$);
		
		\draw[->] (inndec) -- node[above] {$p(w_i|\y_1,\dots,\y_{\numS})$} (outdec);
		\draw[->, dashed] ($(outdec.east) + (0,-0.2)$) -- ($(inndec.west) + (0,-0.2)$);
		\draw[->] (outdec) -- (uh);
		
		\draw [dashed] ($(inn)+(-1.2,1.0)$) rectangle ($(off)+(.9,-.55)$);
		\node[anchor=west] at ($(inn)+(-1.1,.8)$) {Inner code};
		
		\draw [dashed] ($(c1)+(-1.8,1.1)$) rectangle ($(cA)+(.9,-1.1)$);
		\node[anchor=west] at ($(c1)+(-1.8,.9)$) {Channel};
		
	\end{tikzpicture}
	
	\caption{Communication via multiple transmissions over an IDS channel with a concatenated coding scheme. The IDS channel, depicted in Fig.~\ref{fig:ids:channel}, is fed $\numS$ times with the encoded transmit sequence $\x$. }
	\label{fig:system:model}
\end{figure}

The sequence $\x$ is  transmitted $\numS$ times independently over an IDS channel resulting in the received sequences $\y_1,\dots,\y_{\numS}$ corresponding to $\numS$ reads of the original strand. The inner decoder uses these received sequences to infer likelihoods for the symbols in $\w$. These likelihoods are then fed to the outer decoder, which decides on the decoded sequence $\hat{\u}$. Furthermore, we can also iterate between the inner and outer decoder, exchanging extrinsic information between them, which is referred to as \emph{turbo decoding} in the literature.

%
\section{Symbolwise MAP Decoding for the IDS Channel (Inner Decoding)} \label{sec:innerdec-one}
The random memory associated with the insertion and deletion processes  makes decoding for IDS channels challenging. This is visualized by the receiver's inability to directly identify the origin of a received symbol $y_i$. Since insertions or deletions before symbol $y_i$ might have moved the symbol right or left in the received sequence, $y_i$ could be the result of transmitting a symbol $x_{i'}$ with $i'\neq i$. In the following, we describe how it is still possible to infer a posteriori likelihoods from this channel using a hidden Markov representation of the channel \cite{davey_reliable_2001,briffa_improved_2010,buttigieg_improved_2015}. We present and discuss this HMM for $\numS = 1$ first and extend it to $\numS>1$  multiple received sequences afterward. Here we restrict the discussion to convolutional inner codes, as a TVC with $t=1$ can be viewed as a convolutional code with $m=0$.
The APP of the outer code symbol $w_i$ is
\begin{align*}
 p(w_i|\y) = \frac{p(\y,w_i)}{p(\y)}.
 \end{align*}
The joint probability $p(\y,w_i)$ can be computed by de-marginalizing with respect to the memory states of the convolutional code corresponding to symbol $w_i$. This is possible due to the Markov property of the convolutional code. However, this Markov property no longer holds for the IDS channel as a result of its memory. To circumvent this issue, a new hidden state variable, the so-called \emph{drift} \cite{davey_reliable_2001}, that incorporates insertions and deletions, is added to the original Markov process, creating an augmented hidden Markov process. The drift $d_i$, $0 \leq i < \leno + m$, is defined as the number of insertions minus the number of deletions that occurred before symbol $x_{ni+1}$ is enqueued, while $d_{\bl}$ is defined as the number of insertions minus deletions that occurred after the last symbol $x_{n(\leno + m)}$ has been transmitted. Thus, by definition, $d_0 = 0$ and $d_{\bl} = N'-N$, both known to the receiver. In the resulting HMM, a transition from time $i-1$ to time $i$ corresponds to a transmission of symbols $\x_{(i-1)n+1}^{in}$, where $\x_a^b = (x_a,x_{a+1},\dots, x_b)$. Further, when transitioning from state $d_{i-1}$ to $d_{i}$, the HMM emits $n+d_{i}-d_{i-1}$ output symbols depending on both the previous and new drift. The key property of the drift is that, by its inclusion as a new state variable inside the Markov process, the Markov property is restored. This is because the drift sequence itself forms a Markov chain and, conditioned on the memory state and drift at time $i$, the output of the channel after time $i$ becomes independent of the previous memory states and drifts. Introducing the  joint state variable $\sigma_i = (s_i,d_i)$, where $s_i$ denotes the state variables of the convolutional code, we obtain with slight abuse of notation
\begin{align*}
	p(\y,w_i) = \sum_{(\sigma,\sigma'):w_i} p(\y,\sigma,\sigma'),
\end{align*}
where $\sigma$ and $\sigma'$ denote realizations of the random variables $\sigma_{i-1}$ and $\sigma_i$, respectively. The summation is over all convolutional code memory states that correspond to information symbol $w_i$. Using the Markov property, we can expand the joint probability $p(\y,\sigma,\sigma')$ into three parts as
\begin{align*}
	&p(\y,\!\sigma\!,\!\sigma') \!=\!
	p\!\left(\!\y_{1}^{(i\!-\!1)n+d}, \sigma\!\right)\!p\!\left(\!\y_{(i\!-\!1)n\!+d+\!1}^{in+d'}, \sigma'\big|\sigma\!\right)\!p\!\left(\!\y_{in\!+\!d'\!+1}^{\len'}\Big| \sigma'\!\right)\!.
\end{align*}
Abbreviating the above terms by $\alpha_{i-1}(\sigma)$, $\gamma_i(\sigma,\sigma')$, and $\beta_i(\sigma')$ in order of appearance, one can deduce the  forward and backward recursions
\begin{align}
	\alpha_i(\sigma') &= \sum_{\sigma}\alpha_{i-1}(\sigma) \gamma_{i}(\sigma,\sigma'),  \label{eq:alpha_rec}\\
	\beta_{i-1}(\sigma) &= \sum_{\sigma'}\beta_i(\sigma') \gamma_i(\sigma,\sigma'). \notag
\end{align}
Knowing the initial and final drift of the received sequence, the initial and termination conditions of the forward and backward recursions are
\begin{align*}
	\alpha_0(\sigma) &= \begin{cases}
		1& \text{if } \sigma = (0, 0)\\
		0& \text{otherwise},
	\end{cases}
	\\
	\beta_{\leno + m}(\sigma) &= \begin{cases}
		1& \text{if } \sigma = (0, N'-N)\\
		0& \text{otherwise}.
	\end{cases}
\end{align*}
The branch metric can be decomposed as $$\gamma_i(\sigma,\sigma')=p(w_i)p(\y_{(i-1)n+d+1}^{in+d'}, d'\big|d,s,s'),$$ where $p(w_i)$ is the a priori probability of symbol $w_i$. The expression $p(\y_{(i-1)n+d+1}^{in+d'}, d'\big|d,s,s')$ can be efficiently computed using a lattice structure \cite{Buttigieg_ImprovedCC_2014,bahl_decoding_1975} as explained as follows.

\begin{figure}[t]
	\centering
\begin{tikzpicture}[every plot/.style={mark=*, mark size=0.5mm, only marks},scale=0.7]

\draw plot coordinates{(0,0) (0,1) (0,2) (0,3) (0,4) (0,5)
(1,0) (1,1) (1,2) (1,3) (1,4) (1,5)
(2,0) (2,1) (2,2) (2,3) (2,4) (2,5)
(3,0) (3,1) (3,2) (3,3) (3,4) (3,5)
(4,0) (4,1) (4,2) (4,3) (4,4) (4,5)
(5,0) (5,1) (5,2) (5,3) (5,4) (5,5)};

\draw[thick,->,>=stealth] (0,5) -- (1,4);
\draw[thick,->,>=stealth] (1,5) -- (2,4);
\draw[thick,->,>=stealth] (2,5) -- (3,4);
\draw[thick,->,>=stealth] (3,5) -- (4,4);
\draw[thick,->,>=stealth] (4,5) -- (5,4);

\draw[thick,->,>=stealth] (0,5) -- (1,5);
\draw[thick,->,>=stealth] (1,5) -- (2,5);
\draw[thick,->,>=stealth] (2,5) -- (3,5);
\draw[thick,->,>=stealth] (3,5) -- (4,5);
\draw[thick,->,>=stealth] (4,5) -- (5,5);

\draw[thick,->,>=stealth] (0,5) -- (0,4);
\draw[thick,->,>=stealth] (1,5) -- (1,4);
\draw[thick,->,>=stealth] (2,5) -- (2,4);
\draw[thick,->,>=stealth] (3,5) -- (3,4);
\draw[thick,->,>=stealth] (4,5) -- (4,4);
\draw[thick,->,>=stealth] (5,5) -- (5,4);

\draw[thick,->,>=stealth] (0,4) -- (1,3);
\draw[thick,->,>=stealth] (1,4) -- (2,3);
\draw[thick,->,>=stealth] (2,4) -- (3,3);
\draw[thick,->,>=stealth] (3,4) -- (4,3);
\draw[thick,->,>=stealth] (4,4) -- (5,3);

\draw[thick,->,>=stealth] (0,4) -- (1,4);
\draw[thick,->,>=stealth] (1,4) -- (2,4);
\draw[thick,->,>=stealth] (2,4) -- (3,4);
\draw[thick,->,>=stealth] (3,4) -- (4,4);
\draw[thick,->,>=stealth] (4,4) -- (5,4);

\draw[thick,->,>=stealth] (0,4) -- (0,3);
\draw[thick,->,>=stealth] (1,4) -- (1,3);
\draw[thick,->,>=stealth] (2,4) -- (2,3);
\draw[thick,->,>=stealth] (3,4) -- (3,3);
\draw[thick,->,>=stealth] (4,4) -- (4,3);
\draw[thick,->,>=stealth] (5,4) -- (5,3);

\draw[thick,->,>=stealth] (0,3) -- (1,2);
\draw[thick,->,>=stealth] (1,3) -- (2,2);
\draw[thick,->,>=stealth] (2,3) -- (3,2);
\draw[thick,->,>=stealth] (3,3) -- (4,2);
\draw[thick,->,>=stealth] (4,3) -- (5,2);

\draw[thick,->,>=stealth] (0,3) -- (1,3);
\draw[thick,->,>=stealth] (1,3) -- (2,3);
\draw[thick,->,>=stealth] (2,3) -- (3,3);
\draw[thick,->,>=stealth] (3,3) -- (4,3);
\draw[thick,->,>=stealth] (4,3) -- (5,3);

\draw[thick,->,>=stealth] (0,3) -- (0,2);
\draw[thick,->,>=stealth] (1,3) -- (1,2);
\draw[thick,->,>=stealth] (2,3) -- (2,2);
\draw[thick,->,>=stealth] (3,3) -- (3,2);
\draw[thick,->,>=stealth] (4,3) -- (4,2);
\draw[thick,->,>=stealth] (5,3) -- (5,2);

\draw[thick,->,>=stealth] (0,2) -- (1,1);
\draw[thick,->,>=stealth] (1,2) -- (2,1);
\draw[thick,->,>=stealth] (2,2) -- (3,1);
\draw[thick,->,>=stealth] (3,2) -- (4,1);
\draw[thick,->,>=stealth](4,2) -- (5,1);

\draw[thick,->,>=stealth] (0,2) -- (1,2);
\draw[thick,->,>=stealth] (1,2) -- (2,2);
\draw[thick,->,>=stealth] (2,2) -- (3,2);
\draw[thick,->,>=stealth] (3,2) -- (4,2);
\draw[thick,->,>=stealth] (4,2) -- (5,2);

\draw[thick,->,>=stealth] (0,2) -- (0,1);
\draw[thick,->,>=stealth] (1,2) -- (1,1);
\draw[thick,->,>=stealth] (2,2) -- (2,1);
\draw[thick,->,>=stealth] (3,2) -- (3,1);
\draw[thick,->,>=stealth] (4,2) -- (4,1);
\draw[thick,->,>=stealth] (5,2) -- (5,1);

\draw[thick,->,>=stealth] (0,1) -- (1,0);
\draw[thick,->,>=stealth] (1,1) -- (2,0);
\draw[thick,->,>=stealth] (2,1) -- (3,0);
\draw[thick,->,>=stealth] (3,1) -- (4,0);
\draw[thick,->,>=stealth] (4,1) -- (5,0);

\draw[thick,->,>=stealth] (0,1) -- (1,1);
\draw[thick,->,>=stealth] (1,1) -- (2,1);
\draw[thick,->,>=stealth] (2,1) -- (3,1);
\draw[thick,->,>=stealth] (3,1) -- (4,1);
\draw[thick,->,>=stealth] (4,1) -- (5,1);

\draw[thick,->,>=stealth] (0,1) -- (0,0);
\draw[thick,->,>=stealth] (1,1) -- (1,0);
\draw[thick,->,>=stealth] (2,1) -- (2,0);
\draw[thick,->,>=stealth] (3,1) -- (3,0);
\draw[thick,->,>=stealth] (4,1) -- (4,0);
\draw[thick,->,>=stealth] (5,1) -- (5,0);

\draw[above](0,5) node{$l = 0$};
\draw[above](1,5) node{1};
\draw[above](2,5) node{2};
\draw[above](3,5) node{\dots};
\draw[above](4,5) node{$\mu - 1$};
\draw[above](5,5) node{$\mu$};

\draw[left](0,5) node{$r = 0$};
\draw[left](0,4) node{1\;\;\;\;};
\draw[left](0,3) node{2\;\;\;\;};
\draw[left](0,2) node{\vdots\;\;\;\;};
\draw[left](0,1) node{$n - 1$};
\draw[left](0,0) node{$n$\;\;\;\;};

\end{tikzpicture}
\caption{The lattice used to compute $p(\y_{(i-1)n+d+1}^{in+d'}, d'\big|d,s,s')$.}
\label{fig:lattice}
\end{figure}
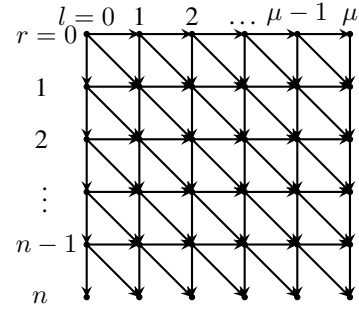

Define the lattice $\boldsymbol{F}_{n \times \mu}$ with $n+1$ rows corresponding to the transmitted sequence $\x_{(i-1)n+1}^{in} = \dot{\x} = (\dot{x}_1,\ldots,\dot{x}_n)$ of length $n$, and $\mu+1 = n + d' - d+1$ columns corresponding to the received sequence $\y_{(i-1)n + d + 1}^{in + d'} = \dot{\y}=(\dot{y}_1,\ldots,\dot{y}_\mu)$ of length $\mu$. A horizontal transition in the lattice represents an insertion  with probability $\nicefrac{p_\I}{q}$, a vertical transition represents a deletion  with probability $p_\D$, while a diagonal transition represents a transmission. The last event has probability $p_\T(1 - p_\S)$ if the corresponding elements in $\dot{\y}$ and $\dot{\x}$ match or probability $\nicefrac{p_\T p_\S}{(q-1)}$ otherwise. Let the value of the lattice point in row $r$ and column $l$ be represented by $F_{r,l}$, see Fig.~\ref{fig:lattice}. Then, for $0 < r < n$ and $0 < l \leq \mu$, a lattice computation is defined recursively as 
\begin{align*}
	F_{r,l} = \frac{1}{q}p_\I F_{r,l-1} + p_\D F_{r-1,l}+ Q(\dot{y_l},\dot{x_r})F_{r-1,l-1},
\end{align*}
where
\begin{align*}
	Q(\dot{y},\dot{x}) = \begin{cases}
		p_\T \frac{p_\S}{q-1}& \text{if } \dot{y} \neq \dot{x}\\
		p_\T (1-p_\S)& \text{otherwise}.
	\end{cases}
\end{align*}
The lattice computation is initialized as
\begin{align*}
	F_{r,l} = \begin{cases}
		1 & \text{if } r = 0, l = 0\\
		0 & \text{otherwise},
	\end{cases}
\end{align*}
and for the last row $(r = n)$, since the HMM does not allow insertions at the end of the transmitted sequence, the lattice computation becomes
\begin{align*}
	F_{n,l} = p_\D F_{n-1,l}+ Q(\dot{y_l},\dot{x_n})F_{n-1,l-1}.
\end{align*}
Finally, $p(\y_{(i-1)n+d+1}^{in+d'}, d'\big|d,s,s') = F_{n,\mu}$. 

\section{Symbolwise MAP Decoding for Multiple Received Sequences} \label{sec:innerdec-multiple}
In this work, we propose two novel approaches to take advantage of the redundancy arising from multiple received sequences. The first approach,
which we refer to as \emph{joint decoding},
calculates the APPs exactly  and works on a multi-sequence state-based trellis (or a joint trellis). The second approach, referred to as \emph{separate decoding}, considers each sequence trellis separately and then combines their respective APPs. Although the first approach offers the best solution for the multiple received sequences  problem, it comes with the drawback of very high complexity as will be shown later. As a result, the second approach is proposed as a more practical and less complex solution to the same problem, at the expense of a slight reduction in performance.
\subsection{Joint Decoding}
To calculate the  APPs of code symbols $w_i$ given multiple received sequences, we proceed as follows. We introduce a drift state vector $\d_i = (d_{i,1},\dots,d_{i,\numS})$ representing the drift at time instant $i$ for each sequence. The resulting HMM has the combined $(\numS+1)$-dimensional state variables \mbox{$\sigma_i = (s_i,d_{i,1},\dots,d_{i,\numS} )$} with corresponding branch metric
\begin{align*}
	&\gamma_i(\sigma,\sigma') = p(w_i) \prod_{j=1}^{\numS} p\left((\y_j)_{(i-1)n+d_{j}+1}^{in+d_{j}'}, d_{j}'\big|d_{j},s,s'\right),
\end{align*}
where $\y_j$, $1 \leq j \leq \numS$, is the $j$-{th} received sequence. This branch metric can be efficiently computed using the lattice implementation as $\gamma_i(\sigma,\sigma') = p(w_i)  \prod_{j=1}^{\numS} F^j_{n,\mu_j}$, where $F^j_{n,\mu_j}$ is the result of the recursive lattice computation based on the substring $(\y_j)_{(i-1)n+d_{j}+1}^{in+d_{j}'}$ of  the $j$-th received sequence $\y_j$ and $\mu_j=n+d'_j-d_j$ is its length. The APPs can then be computed applying the BCJR algorithm  with  initial and termination conditions 

\begin{align*}
	\alpha_0(\sigma) &= \begin{cases}
		1& \text{if } \sigma = (0, 0,\dots,0)\\ 
		0& \text{otherwise},
	\end{cases}
	\\
	\beta_{\leno + m}(\sigma) &= \begin{cases}
		1& \text{if } \sigma = (0, N'_1-N,\ldots,N'_\numS-N)\\
		0& \text{otherwise},
	\end{cases}
\end{align*}
where $N'_j$ denotes the length of the $j$-th received sequence. 
\subsection{Separate Decoding}
We propose a decoding approach that approximates the APPs by decoding each received sequence  separately and combining the resulting single sequence APPs. The proposed separate decoding yields a significant reduction in complexity compared to joint decoding, as discussed in Section~\ref{subsec:complexity}.  More precisely, we approximate the joint APPs $p(w_i|\y_1,\ldots,\y_\numS)$ given the single sequence APPs $p(w_i|\y_j)$, $1 \leq j \leq \numS$, as
\begin{align*}
    p(w_i|\y_1,\dots,\y_{\numS}) \approxprop \frac{\prod_{j=1}^{\numS} p(w_i|\y_j)}{p(w_i)^{\numS-1}}.
\end{align*}
The individual APPs $p(w_i|\y_j)$ can be calculated via the BCJR decoder presented in Section~\ref{sec:innerdec-one}. As we show in Appendix~\ref{app:app:memoryless}, this  approximation holds exactly for memoryless channels, which motivates the proposed separate decoding. However, it results here in a loss of performance, which we  discuss in Sections~\ref{sec:air_BCJRonce} and \ref{sec:simresults}. Note that the separation of the joint APPs could have been carried out as well at other stages in the decoding flow, e.g., directly between the channel and the inner code decoder or after decoding the outer code. However, we observed that for the considered parameters, the proposed stage yields the best results.

\subsection{Complexity Analysis} \label{subsec:complexity}
In this subsection, we  evaluate the  complexity of joint and separate decoding. The number of performed operations by the inner BCJR decoding is mainly proportional to the number of edges in the trellis inferred by the HMM. In line with \cite{davey_reliable_2001}, for all methods we limit the drift  to a fixed interval, $d_{i,j} \in [d_\mathrm{min},d_{\max}]$, and the maximum number of insertions per symbol to $I_\mathrm{max}$. Let us denote by $\Delta = d_\mathrm{max}-d_{\min}+1$ the total number of drift states. Moreover, we can quantify the total number of possible drift transisitions as $\delta = n(I_\mathrm{max}+1)+1$. Let $\nu$ be the number of binary memory elements of the convolutional encoder. The complexity to decode a single received sequence by the inner BCJR deocoder is
\begin{align*}
    C_{\mathrm{single}} = \frac{N}{n} 2^{\nu+k}\Delta \delta.
\end{align*}
When considering the multiple sequence case, with $\numS$ being the total number of sequences, the complexity of the presented methods can be computed as follows. For the separate decoding, the complexity is simply $\numS$ times that of the single sequence decoding, i.e., 
\begin{align*}
    C_{\mathrm{sep}} = \numS \cdot C_{\mathrm{single}} = \frac{N}{n} 2^{\nu+k} \numS \Delta \delta.
\end{align*}
Since the joint decoding introduces a joint drift state vector, the decoding trellis grows exponentially in the number of sequences $\numS$. Thus, the complexity of joint decoding is
\begin{align*}
    C_{\mathrm{joint}} = \frac{N}{n} 2^{\nu+k}(\Delta \delta)^\numS.
\end{align*}
Note that we do not take into account the effect of the trellis termination on the number of edges in the aforementioned decoding methods. Moreover, typical values for the product $\Delta \delta$ can be as high as $ \approx 1000$ in the DNA storage application. Therefore, the inner code decoding is the main contributor to the overall complexity and thus we have neglected the complexity of the outer code decoding. 

For a fixed length $\len$, increasing the number of sequences $\numS$ has different impact in the scale of complexity. 
The complexity of the separate decoding method scales linearly with $\numS$ while the complexity of the joint decoding approach scales exponentially with $\numS$. Therefore, it directly becomes evident that the joint decoding method is only practical for a small number of sequences. Hence, the reason we proposed the separate decoding algorithm. With separate decoding, one can increase the number of sequences $M$ in order to match the performance of optimal joint decoding with a significantly lower complexity. As illustrated in later sections of the paper, the number of sequences needed to match the performance of joint decoding with $M=2$ is low.


%
\section{Concatenated Code Design} \label{sec:concatenated_code_design}

In this section, we discuss the optimization of the coding scheme. On a high level, we proceed as follows. We first design a novel inner TVC that is optimized in terms of the Levenshtein distance between two codewords in a block and also avoids overlaps between the codebooks of two adjacent blocks, improving its time variance. We evaluate the performance of this inner code by computing AIRs in Section~\ref{sec:air} and compare with codes from the literature. For a selected inner code, we then proceed with optimizing an outer LDPC code using protographs and density evolution (DE). We also optimize the frozen symbols of an outer nonbinary polar code using Monte-Carlo methods.
\subsection{Inner Block Code} \label{sec:inner:code}
 Our goal is to improve on the performance of the inner code construction introduced in \cite{davey_reliable_2001}, which we refer to as the Davey-MacKay (DM) construction. The authors of \cite{davey_reliable_2001} proposed to construct a binary inner block code of size $2^k$ and length $n$ by selecting the $2^k$ vectors of lowest Hamming weight from all $2^n$ vectors of length $n$. In other words, the  constructed code consists of the $2^k$ sparsest length-$n$ binary vectors. We refer to the combination of the block code arising from the DM construction and a random offset sequence as a \emph{watermark}  code. 

The use of a random offset sequence is essential as it helps the inner code in maintaining synchronization with the transmitted sequence. More precisely, the random sequence helps in tracking the codeword boundaries between consecutive transmissions of inner code codewords. This in turn helps the inner code to avoid confusing adjacent codewords, especially in the presence of a large number of insertions and deletions. The authors in \cite{davey_reliable_2001} state that the more random the watermark code is, the better the synchronization, which motivated their sparse code construction, as it alters little the random sequence. However, for low IDS probabilities, where loss of synchronization is less probable, the dominating factor of the inner code performance is its \emph{edit distance} profile. The edit distance between two codewords, denoted by $d_{\mathsf{e}}$, is defined as the minimum number of  IDS errors required to change one codeword into another. Hence, an inner code with good minimum $d_{\mathsf{e}}$,  $d_{\mathsf{e}}^{\text{min}}$, will perform well in that region.  We observe this phenomenon in the AIR results in Section~\ref{sec:air} (for an in-depth discussion on the synchronization ability and performance of inner codes, we refer the reader to that section). In the special case when no substitutions are allowed, the edit distance is typically  referred to as the Levenshtein distance, which we denote by $d_{\mathsf{L}}$.     

It was shown in \cite{Briffa_ImprovementOT_2008} that the DM construction is far from optimal due to its poor Levenshtein distance profile. The authors in \cite{Nguyen_OnTW_2013} considered the so-called \emph{weighted} Levenshtein distance and improved its  profile for the watermark code, hence  the code performance, by first building a set of sequences that when added to the codewords, improve the overall distance profile of the code. Secondly, for every transmission, a sequence is randomly picked from this  set. However, the performance is still relatively poor due to a low $d_{\mathsf{e}}^{\text{min}}$. In \cite{Buttigieg_CodebookAM_2011,Buttigieg_ImprovedCC_2014}, it was proposed to use codewords from Varshamov-Tenegol'ts (VT) codes \cite{VTcodes1965}, known for their good IDS correction capabilities, to construct an inner code. In particular, in \cite{Buttigieg_ImprovedCC_2014}, the authors chose a set of codewords from a VT code using a simulated annealing algorithm \cite{Gamal1987UsingSA} that targets the minimization of the so-called change probability of the set. One more issue to combat comes from the requirement of using a random offset sequence which does not destroy the  edit distance properties of the inner code. This issue was previously addressed by the same authors in \cite{Buttigieg_CodebookAM_2011} where they  proposed to  randomly choose an offset sequence  only from the set of sequences that when added to the codewords of the inner code do not alter its $d_{\mathsf{e}}^{\text{min}}$.\footnote{Note that in \cite{Buttigieg_CodebookAM_2011} the edit distance is referred to as the Levenshtein distance.}  However, such inner codes perform better when the codeword boundaries between consecutive transmissions are known, which is typically not the case. As a result, in the later work \cite{Buttigieg_ImprovedCC_2014}, TVCs, which are composed of several alternating codebooks found by minimizing the change probability  through a simulated annealing search, were proposed. One issue that may arise is that using the simulated annealing algorithm does not guarantee all codebooks to have the best $d_{\mathsf{e}}^{\text{min}}$.

\begin{table} 
	\setlength{\tabcolsep}{2pt}
		\caption{Designed TVC With $4$ Codebooks}
		\vspace{-2ex}
		\label{tab:codebooks}
	\begin{center}
		
		{\renewcommand{\arraystretch}{1.1}
			\begin{tabular}{c|c|c|c} \specialrule{1.2pt}{0pt}{0pt}
				Codebook $1$ & Codebook $2$ & Codebook $3$ & Codebook $4$\\ \specialrule{.8pt}{0pt}{0pt}
				$(0,0,0,0)$ & $(0,0,0,1)$ & $(0,0,3,0)$ & $(0,0,0,3)$\\

				$(0,0,2,2)$ & $(0,0,3,3)$ & $(0,1,1,1)$ & $(0,0,2,2)$\\

				$(0,3,2,3)$ & $(0,2,1,2)$ & $(0,2,3,2)$ & $(0,3,2,3)$\\

				$(1,0,1,0)$ & $(1,0,2,0)$ & $(0,3,1,3)$ & $(1,0,1,1)$\\

				$(1,1,1,1)$ & $(1,1,2,2)$ & $(1,0,0,1)$ & $(1,1,0,0)$\\

				$(1,1,3,3)$ & $(1,1,3,1)$ & $(1,2,0,2)$ & $(1,1,3,3)$\\

				$(1,2,3,2)$ & $(1,3,0,3)$ & $(1,3,2,3)$ & $(1,2,3,2)$\\

				$(2,0,2,1)$ & $(2,0,0,2)$ & $(2,2,0,0)$ & $(2,1,2,1)$\\

				$(2,1,2,0)$ & $(2,2,0,3)$ & $(2,2,1,3)$ & $(2,2,0,0)$\\

				$(2,2,2,2)$ & $(2,2,1,1)$ & $(2,2,2,2)$ & $(2,2,2,3)$\\

				$(2,2,3,3)$ & $(2,3,1,3)$ & $(2,3,0,3)$ & $(2,3,0,3)$\\

				$(3,0,3,1)$ & $(3,0,1,0)$ & $(3,0,0,2)$ & $(3,0,0,1)$\\

				$(3,1,3,0)$ & $(3,2,2,2)$ & $(3,2,1,2)$ & $(3,1,3,0)$\\

				$(3,2,0,0)$ & $(3,2,3,0)$ & $(3,3,1,1)$ & $(3,2,0,2)$\\

				$(3,3,2,2)$ & $(3,3,1,1)$ & $(3,3,2,0)$ & $(3,2,3,1)$\\

				$(3,3,3,3)$ & $(3,3,3,2)$ & $(3,3,3,3)$ & $(3,3,3,3)$\\
                \specialrule{.8pt}{0pt}{0pt}
			\end{tabular}}
	\end{center}
	\vspace{-5ex}
\end{table}

Here, we consider nonbinary TVCs, where the codebooks are found by searching for cliques in an undirected graph as described below. Searching for codebooks in this way guarantees that they all have the same $d_{\mathsf{e}}^{\text{min}}$.
First,  an undirected graph with a vertex for each vector in $\Sigma_{q}^n$ is constructed. Then, we search for a maximum clique in that graph. In general, finding a maximum clique in a graph is an NP-hard problem. However, for small graphs, there exist several efficient algorithms that can be used \cite{stergrd2002clique1, Sewell1998clique2, Carraghan1990clique3}. Here, we use the  branch-and-bound algorithm proposed in  \cite{Sewell1998clique2}. Formally, let $\mathcal{G} = (\mathcal{V}, \mathcal{E})$ be an undirected graph, where $\mathcal{V}$ denotes  the set of vertices and $\mathcal{E}$ the set of edges. Each vertex $v=v(\boldsymbol{c}) \in \mathcal{V}$ represents a nonbinary vector $\boldsymbol{c} \in \Sigma_{q}^n$, and there is an edge between vertices $v(\boldsymbol{c}_i)$ and $v(\boldsymbol{c}_j)$ if and only if $d_{\mathsf{e}}(\boldsymbol{c}_i, \boldsymbol{c}_j) \geq d_{\mathsf{e}}^{\text{min}}$, where $d_{\mathsf{e}}^{\text{min}}$ is a prescribed minimum edit distance.\footnote{Here, $d_{\mathsf{e}}(\cdot,\cdot)$ denotes the edit distance between its first and second argument. Moreover, when computing the distance, according to the IDS channel model, we do not allow for insertions at the end.} Then, an $[n,k]_q$ (nonlinear) block code with edit distance at least equal to $d_{\mathsf{e}}^{\text{min}}$ corresponds to a clique of size $2^k$ in $\mathcal{G}$.

The branch-and-bound algorithm searches for a set of cliques (or codebooks) of a given size, and from this set we select $t$ codebooks that overlap as little as possible  using a heuristic approach. The $t$ selected codebooks can then either be repeated periodically or randomly in order to construct a TVC. In this work, we only consider the special case of no substitutions when constructing a TVC. Hence, we consider the  Levenshtein distance. 
In Table~\ref{tab:codebooks}, we list $t=4$ quaternary codebooks, each of size $16$, length $n=4$, and with minimum Levenshtein distance $d_{\mathsf{L}}^{\text{min}} = 4$.  Note that for these code parameters, finding $4$ disjoint codebooks is difficult. Hence, some of the codebooks overlap and the number of distinct codewords is $56$. 
The resulting TVC is a $[4,4,4]_{4}$ TVC.

In Section~\ref{sec:simresults}, we compare the performance of our concatenated coding schemes using a quaternary watermark code and the quaternary TVC code in Table~\ref{tab:codebooks} as inner codes. We further consider the use of  a convolutional inner code. Here, we do not extend the inner code optimization to the convolutional code, but rather pick the same code as in \cite{Buttigieg_ImprovedCC_2014}, which is the one corresponding to the generator polynomial $g = [5, 7]_{\text{oct}}$. By grouping trellis sections together in the decoding stage, we can interpret this convolutional code also as a nonbinary inner code, thus forming APPs for higher order field sizes. 

\subsection{Outer Code: Low-Density Parity-Check Code} \label{sec:outerLDPC}

We consider protograph LDPC codes as they facilitate achieving lower error floors compared to unstructured codes. Formally,  a  protograph is a small  multi-edge-type graph with $n_{\mathsf{p}}$ variable-node (VN) types and $r_{\mathsf{p}}$ check-node (CN) types. A protograph can be represented by a base matrix 
\begin{align*}
    \boldsymbol{B} = \begin{pmatrix}
    b_{0,0} & b_{0,1} & \dots & b_{0,n_{\mathsf{p}}-1} \\
    b_{1,0} & b_{1,1} & \dots & b_{1,n_{\mathsf{p}}-1} \\
    \vdots & \vdots & \dots & \vdots \\
     b_{r_{\mathsf{p}}-1,0} & b_{r_{\mathsf{p}}-1,1} & \dots & b_{r_{\mathsf{p}}-1,n_{\mathsf{p}}-1}\\
    \end{pmatrix},
\end{align*}
where  entry $b_{i,j}$ is an integer representing the number of edge connections from a type-$i$ VN to a type-$j$ CN. A parity-check matrix $\boldsymbol{H}$ of an LDPC code can then be constructed by lifting the base matrix $\boldsymbol{B}$ by replacing each nonzero (zero) $b_{i,j}$ with a $Q_{\mathsf{p}} \times Q_{\mathsf{p}}$ circulant (zero) matrix with  row and column weight equal to $b_{i,j}$.  The circulant matrices are picked in order to maximize the girth of the corresponding Tanner graph by using  the progressive edge-growth algorithm~\cite{PEG}. The resulting lifted   parity-check matrix, of dimensions $Q_{\mathsf{p}}r_{\mathsf{p}} \times Q_{\mathsf{p}}n_{\mathsf{p}}$,  defines an LDPC code of length $Q_{\mathsf{p}}  n_{\mathsf{p}}$ and dimension at least $Q_{\mathsf{p}}(n_{\mathsf{p}}-r_{\mathsf{p}})$. To construct a nonbinary code from the lifted matrix, we randomly assign nonzero entries from $\field{\outq}$ to the edges of the corresponding Tanner graph.

We optimize the protograph using DE.  

Particularly, we consider the DE algorithm proposed in \cite{Kavcic2003BinaryII} for the optimization of binary LDPC codes
for intersymbol interference channels, extended to nonbinary protographs. The algorithm is based on estimating the probability density functions of the messages from the inner code to the outer LDPC code via Monte-Carlo simulations of the inner decoder and channel detector.  When performing the DE, we assume uniformly random protograph edge weights. The optimized protograph is then chosen as the one yielding the best iterative decoding threshold, $p_\mathsf{th}$. The iterative decoding threshold  predicts the asymptotic performance of the code, in the sense that the bit error probability under iterative decoding approaches zero for $p_\I = p_\D < p_{\mathsf{th}}$, for a given $p_{\S}$, as the block length goes to infinity. The optimization problem is hence
  \begin{align*}
	\underset{\boldsymbol{B}}{\arg\max} \quad &p_{\mathsf{th}}\\
	\text{s.t.} \quad \quad &b_{i,j} \leq b_{\text{max}}, \quad \forall\, (i,j),\\ 
	\quad \quad &n_{\mathsf{p}} \leq n^{\text{max}}_{\mathsf{p}},\\
	&\frac{n_{\mathsf{p}}-r_{\mathsf{p}}}{n_{\mathsf{p}}} = \rateo,
\end{align*}
where $b_{\text{max}}$ is the maximum allowed circulant weight, $n^{\text{max}}_{\mathsf{p}}$ is the maximum allowed number of different VN types, and $\rateo$ is the design rate of the LDPC code. 
  
One issue with optimizing over protographs is that neither the dimensions of the protograph  nor the number of edge connections have a natural limit. To make the search feasible, we restrict the search to $n^{\text{max}}_{\mathsf{p}}=4$ and $b_{\text{max}}=2$.

\subsection{Outer Code: Polar Code} \label{sec:outerPolar}
As another choice for the nonbinary outer code, we consider the polar code construction from \cite{SteinerYuan2019_NonBinaryPolarCodes} over the field \field{\outq}. The construction's main feature is the extension of Ar{\i}kan's binary kernel \cite{Arikan2009_Polarcode} to a nonbinary kernel 
\begin{align*}
    \kernel = \begin{pmatrix}
    1 & 0 \\
    \alpha  & \beta \\
    \end{pmatrix},
\end{align*}
where $\alpha, \beta \in \field{\outq}$. The encoding can be written as $\ocw = \oinffrz \kernel ^{\otimes \log_{\outq} \olen}$, where $(\cdot)^{\otimes}$ denotes the  Kronecker power and $\oinffrz \in \vecspace{\outq}{\olen}$. One can determine a mapping from the actual information vector $\oinfnofrz \in \vecspace{\outq}{\odim}$ to $\oinffrz$ by fixing $\olen - \odim$ positions, denoted as \emph{frozen positions}, to a specific symbol and padding the vector $\oinfnofrz$ into the remaining positions. For decoding, we use  successive cancellation list (SCL) decoding over $\field{\outq}$ as described in \cite{SteinerYuan2019_NonBinaryPolarCodes,TalVardy_ListDecPolar_2015}. To boost the decoding performance of the SCL decoder, a cyclic redundancy check (CRC) of bit length $\crclen$ is applied. To ensure comparability for fixed rates, the number of frozen positions is hence reduced to $\olen - \odim - \nicefrac{\crclen}{\log(\outq)}$.\footnote{ To simplify notation, in the rest of the paper, $\log$ denotes  logarithm to the base $2$.} 

In this work, we optimize two  components of the nonbinary polar code for our channel using Monte-Carlo simulations: the kernel selection and the frozen positions. For the kernel, we optimize the single-level polarization effect as described in \cite{SteinerYuan2019_NonBinaryPolarCodes}, which means that the optimization step is only done via considering a single $2 \times 2$ kernel. We revise the method shortly here while directly applying our channel model. We treat the inner code including the IDS channels as an auxiliary channel. For this channel, we generate random input sequences $\ocw$ uniformly at random and transmit them via the auxiliary channel to gain APP samples $p(w_i | \chout)$, where $\chout = (\chout_1, \ldots, \chout_\numseq)$. 

First, fix $\alpha, \beta \in \field{\outq}$ and do the following. 
\begin{enumerate}
    \item Choose an input vector $\oinffrz = (u_1, u_2) = (0, u_2)$, where $u_2 \in \field{\outq}$ is chosen uniformly at random. 
    \item Calculate $\ocw = (w_1, w_2) = (u_1 + \alpha u_2, \beta u_2)$.
    \item Pick uniformly at random two APP samples $p(w_1 | \chout)$ and $p(w_2 | \chout)$  under the constraint that these correspond to the values in the vector $\ocw$.
    \item Calculate $p(u_2 | u_1=0, \chout)$ using the APP samples.
\end{enumerate}
Consequently, we have that $p(u_2| u_1=0, \chout)$ is a random variable depending only on the auxiliary channel. 
Hence, we can find  the best ratio $\nicefrac{\alpha}{\beta}$ via  Monte-Carlo simulations by choosing 
\begin{align*}
    \frac{\alpha}{\beta} = \underset{\frac{\alpha}{\beta} \in \field{\outq}}{\arg \min} \; \mathbb{E} \left(1- p(u_2| u_1=0, \chout) \right),
\end{align*}
where the expectation is taken with respect to the random variable on the auxiliary channel with fixed inner code and channel parameters. 

\begin{figure}[t]
    \centering
            \begin{tikzpicture}
        \begin{semilogyaxis}[
                legend style={nodes={scale=0.7, transform shape}},
                width = 0.95*\columnwidth,
                xlabel = {$\nicefrac{\alpha}{\beta}$},
                xlabel style = {nodes={scale=0.7, transform shape}},
,                ylabel = {$\mathbb{E}\left(1- p(u_2| u_1=0, \chout)\right) $},
                xmin = 0,
                xmax = 16,
                xtick distance=1,
                ymin = 0.000001,
                ymax = 1,
                legend style = {at={(0.15,0.9)},anchor=west},
                legend cell align=left,
                legend columns=3, 
                grid=both,
                grid style={solid, gray!30},
                cycle list name=color list]

\addplot+ [solid, mark=triangle] table {
1.000000 0.000501
2.000000 0.000019
3.000000 0.000009
4.000000 0.000022
5.000000 0.000013
6.000000 0.000013
7.000000 0.000013
8.000000 0.000020
9.000000 0.000023
10.000000 0.000012
11.000000 0.000009
12.000000 0.000011
13.000000 0.000025
14.000000 0.000011
15.000000 0.000026
};
\addlegendentry{$p=0.01$ };

\addplot+ [solid, mark=o] table {
1.000000 0.001388
2.000000 0.000196
3.000000 0.000127
4.000000 0.000173
5.000000 0.000103
6.000000 0.000109
7.000000 0.000118
8.000000 0.000181
9.000000 0.000198
10.000000 0.000127
11.000000 0.000112
12.000000 0.000108
13.000000 0.000186
14.000000 0.000139
15.000000 0.000178
};
\addlegendentry{$p=0.02$ };

\addplot+ [solid, mark=square] table {
1.000000 0.002911
2.000000 0.000682
3.000000 0.000472
4.000000 0.000618
5.000000 0.000422
6.000000 0.000468
7.000000 0.000452
8.000000 0.000590
9.000000 0.000699
10.000000 0.000467
11.000000 0.000419
12.000000 0.000437
13.000000 0.000603
14.000000 0.000466
15.000000 0.000610
};
\addlegendentry{$p=0.03$ };

\addplot+ [solid, thick, mark=diamond] table {
1.000000 0.005475
2.000000 0.001937
3.000000 0.001463
4.000000 0.001812
5.000000 0.001369
6.000000 0.001433
7.000000 0.001462
8.000000 0.001779
9.000000 0.001978
10.000000 0.001491
11.000000 0.001367
12.000000 0.001465
13.000000 0.001788
14.000000 0.001424
15.000000 0.001762
};
\addlegendentry{$p=0.04$ };

\addplot+ [solid, mark=pentagon] table {
1.000000 0.016951
2.000000 0.009787
3.000000 0.008436
4.000000 0.009585
5.000000 0.008195
6.000000 0.008316
7.000000 0.008354
8.000000 0.009286
9.000000 0.009884
10.000000 0.008411
11.000000 0.008299
12.000000 0.008441
13.000000 0.009504
14.000000 0.008363
15.000000 0.009258
};
\addlegendentry{$p=0.06$ };

\addplot+ [solid, mark=star] table {
1.000000 0.044320
2.000000 0.033034
3.000000 0.030062
4.000000 0.032158
5.000000 0.029714
6.000000 0.029818
7.000000 0.029798
8.000000 0.031890
9.000000 0.032920
10.000000 0.030147
11.000000 0.029851
12.000000 0.030118
13.000000 0.032351
14.000000 0.029938
15.000000 0.031880
};
\addlegendentry{$p=0.08$ };

        \end{semilogyaxis}
        \end{tikzpicture}
    \vspace{-3ex} 
    \caption{Monte-Carlo simulations of $\mathbb{E} \left( 1- p(u_2| u_1=0, \chout) \right)$ for $\numS = 1$, $p_{\S}=0$, and different $p_{\I}=p_{\D}=p$ on different ratios $\alpha / \beta$ with fixed $\beta =1$, $\outq = 16$, and primitive polynomial $x^4+x+1$. The inner code is a convolutional code with generator polynomial $g = [5,7]_\text{oct}$ with a random offset sequence.}
    \label{fig:polar-kernel}
\end{figure}
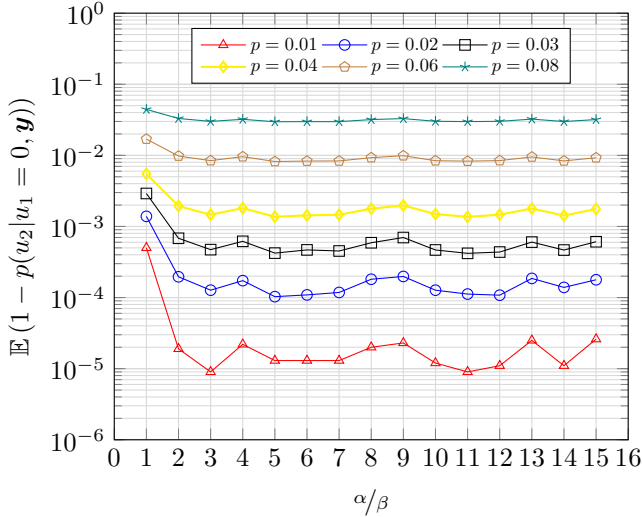

This optimization is dependent on the choice of the inner code and may vary for different codes. In Fig.~\ref{fig:polar-kernel}, the failure rate $\mathbb{E} \left(1- p(u_2| u_1=0, \chout) \right)$ is depicted as a function of the ratio $\nicefrac{\alpha}{\beta}$ for various insertion and deletion probabilities  $p_{\I}=p_{\D}=p$, with $p_{\S}=0$ and $\numS=1$. We denote the field elements $\alpha, \beta \in \mathbb{F}_{\outq}$ as integers such that their binary transformation corresponds to the coefficients of the respective polynomial representation. Since the goal is to find the best kernel selection, i.e., minimize the failure rate over all possible values $p$ for a given inner code, good choices of the kernel parameters for the depicted setup would be $\alpha = 3,5,10,11,12$ while fixing $\beta = 1$.

Moreover, we determine the frozen positions via a genie-aided rate-one polar code with a fixed inner coding scheme by using Monte-Carlo simulations.

%
\section{Achievable Information Rates} \label{sec:air}
We now turn to presenting AIRs over multiple IDS channels to benchmark our coding schemes. We follow two standard approaches for the estimation of AIRs over hidden Markov channels, i.e., channels whose output is the output of a HMM. The first approach is to compute the so-called \emph{BCJR-once} rate \cite{Kavcic2003BinaryII,muller_capacity_2004,soriaga_determining_2007}, which is defined as the symbolwise mutual information between the input process and its corresponding log-likelihood ratios, produced by a symbolwise MAP detector (BCJR algorithm). The second approach is based on \cite{pfister_achievable_2001,arnold_simulation-based_2006} and uses concentration properties of Markov chains to estimate the mutual information between channel input and output. 

\subsection{BCJR-Once Rate} \label{sec:air_BCJRonce}
The BCJR-once rate \cite{Kavcic2003BinaryII,muller_capacity_2004,soriaga_determining_2007}, which we denote by $R_{\text{BCJR-once}}$,  serves as an information rate that can be achieved using an inner decoder that passes once symbolwise APPs to an appropriate outer decoder that is unaware of possible correlations between the symbolwise estimates. The outer and inner decoders hereby perform no iterations. The BCJR-once rate can be derived by computing an AIR of a mismatched decoder as follows. While computing achievable rates for mismatched decoders is well understood \cite{szczecinski_bit-interleaved_2015}, we shortly review the most important derivations for completeness and convenience of the readers. To this end, denote by $q(\w|\y)$, where $\y = (\y_1,\dots,\y_\numS)$, an arbitrary decoding metric that is a valid distribution, i.e., $\sum_{\w} q(\w|\y) = 1$ for all $\y$ and satisfies $q(\w|\y) = 0$ if $p(\w|\y) = 0$. In the following, let $H(\cdot)$ denote the entropy function. We obtain for the mutual information between the message $\w$ and output $\y$,\footnote{With some abuse of notation, for simplicity, we do not distinguish notationwise between random variables and their realizations. Hence, e.g., $\w$ in $H(\w)$ denotes a random vector, while $p(\w)$ denotes the probability of a given realization.}
\begin{align}
	I(\w;\y) &= H(\w) - H(\w|\y) \nonumber \\
	&= H(\w) + \sum_{\w,\y} p(\w,\y) \log p(\w|\y) \nonumber \\
	&= H(\w) + \sum_{\w,\y} p(\w,\y) \left(\log q(\w|\y) + \log\frac{p(\w|\y)}{q(\w|\y)}\right) \nonumber \\
	&\geq H(\w) + \sum_{\w,\y} p(\w,\y) \log q(\w|\y), \label{eq:mi:mismatched}
\end{align}
where the last inequality is due to identifying the sum over the second summand as a Kullback-Leibler divergence, which is nonnegative. Next, we concretize the mismatched metrics for the BCJR-once rate and separate decoding, including a description of how we numerically estimate the above terms. Using the mismatched decoding setup, we can obtain the BCJR-once rate by computing the AIR of a mismatched decoder with decoding metric
$$q_\text{BCJR}(\w|\y) = \prod_{i=1}^{\leno + m} q(w_i|\y).$$
Note that here also the symbolwise a posteriori likelihoods $q(w_i|\y)$ are considered mismatched, due to the fact that their trellis-based computation is not exact as we truncate some of the edges and states for complexity reasons. For the case of decoding multiple received sequences with separate decoding, we use the mismatched metric
$$ q_\text{BCJR-sep}(\w|\y_1,\dots,\y_{\numS}) \propto \prod_{i=1}^{\leno + m}\prod_{j=1}^{\numS} q(w_i|\y_j), $$
where the proportionality constant is chosen such that the decoding metric is a valid distribution, i.e., $\sum_{\w} q_\text{BCJR-sep}(\w|\y_1,\dots,\y_{\numS}) = 1$. Defining the associated mismatched log-likelihood ratios
$$ L^{\text{BCJR-sep}}_i(a) = \sum_{j=1}^{\numS} \ln \frac{q(w_i=a|\y_j)}{q(w_i=0|\y_j)} $$
for all $a \in \field{\outq}$, we can combine the mismatched metric with the mismatched log-likelihood ratios to obtain
$$ q_\text{BCJR-sep}(\w|\y_1,\dots,\y_{\numS}) = \prod_{i=1}^{\leno + m} \frac{\e^{L^{\text{BCJR-sep}}_i(w_i)}}{\sum_{a \in \field{\outq}}\e^{L^{\text{BCJR-sep}}_i(a)}},$$
where the denominator has been chosen such that we obtain a valid distribution. Plugging the result into \eqref{eq:mi:mismatched} yields
\begin{align*}
	I(\w;\y) &\geq H(\w) + \sum_{\w,\y} p(\w,\y) \log q_\text{BCJR-sep}(\w|\y) \\
	&=H(\w) + \sum_{i=1}^{\leno + m} \sum_{w_i,\y} p(w_i,\y) \log\frac{\e^{L^{\text{BCJR-sep}}_i(w_i)}}{\sum_{a \in \field{\outq}}\e^{L^{\text{BCJR-sep}}_i(a)}}.
\end{align*}
For independent and uniform inputs, we obtain $H(\w) = (\leno + m) \log \outq$. Under the assumption that the expectation above obeys asymptotic ergodicity, we conclude that, for large $\leno$, we can estimate the BCJR-once rate by sampling a long input sequence $\w$ and corresponding output sequence $\y$ and compute the log-likelihood ratios $L^{\text{BCJR-sep}}_i(a)$, allowing to conclude with the estimate
$$R_{\text{BCJR-once}} \approx \ratei \log \outq + \frac{\ratei}{\leno + m} \sum_{i=1}^{\leno + m} \log\frac{\e^{L^{\text{BCJR-sep}}_i(w_i)}}{\sum_{a \in \field{\outq}}\e^{L^{\text{BCJR-sep}}_i(a)}}.$$
Note that in the above expression, we have taken into account that the BCJR-once rate is measured in terms of information bits per channel use and we thus normalize the above sum with respect to the number of channel uses $\nicefrac{(\leno + m)}{\ratei}$. This expression has previously been derived in, e.g., \cite{szczecinski_bit-interleaved_2015,hagenauer_exit_2004} and is also valid for mismatched log-likelihood ratios, as we have shown in our derivation. It is important to mention that to compute this expression, it is necessary to use both the mismatched log-likelihood ratios $L^{\text{BCJR-sep}}_i(a)$ and the original input sequence $\w$. Other approaches to compute the BCJR-once rate \cite{land_computation_2004,Kliewer2006} rely on the correctness of the computed log-likelihood ratios. In that case, it is possible to estimate the rate using a formula that only depends on the log-likelihood ratios and \emph{not} the input sequence $\w$, as pointed out in \cite{hagenauer_exit_2004}. However, in our case we compute mismatched log-likelihood ratios, and thus such an approach might give incorrect results.
\begin{table}[t]
	\setlength{\tabcolsep}{1.7pt}
	\begin{center}
		\caption{Inner Code Scheme Selection}
		\vspace{-3ex}
		{\renewcommand{\arraystretch}{1.1}
			\begin{tabular}{clccc} \specialrule{1.2pt}{0pt}{0pt}
				Scheme & Inner code & Gen. polynomial & Alt. pattern & Rate \\ \specialrule{.8pt}{0pt}{0pt}
				CC-$1$ & $(2,2,2)_4$ Conv. code with RS & $g_1*$ & - & $0.98$ \\
				CC-$2$ & $(4,4,2)_4$ Conv. code with RS & $g_2*$ & - & $0.98$ \\
				WM & $[4,4,1]_4$ Watermark code & - & - & $1.0$ \\
				TVC-$1$ & $[4,4,4]_4$ TVC & - & Random* & $1.0$ \\ 
				TVC-$2$ & $[4,4,4]_4$ TVC with RS & - & CB1 to CB4* & $1.0$ \\
				\specialrule{1.2pt}{0pt}{4pt}
			\end{tabular}\label{tab:Inner code selection}}
	\end{center}
	*The alternating pattern of the TVC-$1$ scheme is done by choosing randomly the $4$ codebooks, denoted by CB1-CB4, from Table~\ref{tab:codebooks} and avoiding consecutive codebooks. For the TVC-$2$ scheme, it is simply done by repeating CB1 to CB4 in a round Robin fashion. The generator polynomials are $g_1=[5,7,12,16]_\text{oct}$ and $g_2=[24,34,70,120,160,240,340]_\text{oct}$, where $g_1$ and $g_2$ correspond to combining $2$ and $4$ consecutive trellis sections of $g = [5,7]_\text{oct}$, respectively. RS is shorthand for random sequence. 
	\vspace{-.4cm}
\end{table}
 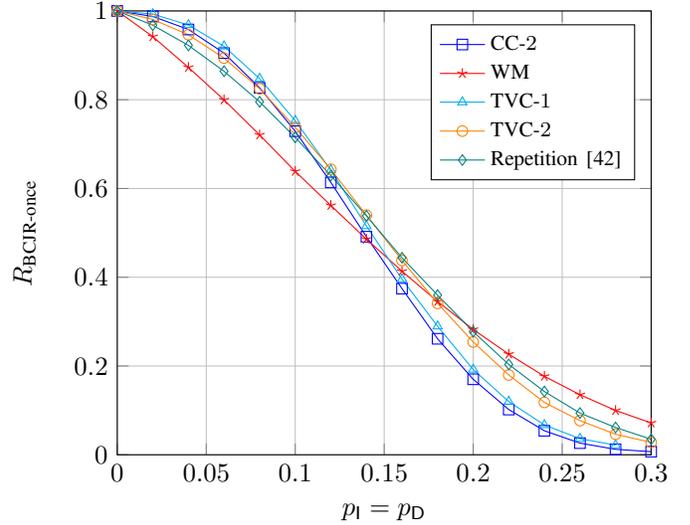
\begin{figure}
 \centering \centering

	\begin{tikzpicture}
		\begin{axis}[
			width = 0.98\columnwidth,
			xmin=0.0,   xmax=0.3,
			ymin=0,	ymax=1,
			xtick distance = 0.05,
			xticklabel style = {/pgf/number format/fixed, /pgf/number format/precision=2},
			grid = both,
			grid style = {line width=.1pt},
			legend pos=north east,
			legend cell align={left},
			legend style={font=\footnotesize},
			xlabel = {$p_\I=p_\D$},
			ylabel style = {name=AIR,yshift=-.0cm},
			ylabel = {$R_{\text{BCJR-once}}$},
			]
			
			\addplot [color=blue,mark=square,each nth point=2] table [col sep=space] {Figures/BCJR-Once/BO_CC4_Off_M1.txt};
			\addlegendentry{CC-$2$};
			
			\addplot [color=red,mark=star,each nth point=2] table [col sep=space] {Figures/BCJR-Once/BO_WM4_Off_M1.txt};
			\addlegendentry{WM};
			
			
			\addplot [color=cyan,mark=triangle,mark options={solid},each nth point=2] table [col sep=space] {Figures/BCJR-Once/BO_RSTV4_NoOff_M1.txt};
			\addlegendentry{TVC-$1$};
			
			\addplot [color=orange,mark=o,each nth point=2] table [col sep=space] {Figures/BCJR-Once/BO_TV4_Off_M1.txt};
			\addlegendentry{TVC-$2$};
			
			\addplot [color= teal,mark=diamond,each nth point=2] table [col sep=space] {Figures/BCJR-Once/BO_Rep4_Off_M1.txt};
			\addlegendentry{Repetition \cite{Srinivasavaradhan2021TrellisBMA}};
		\end{axis}
	\end{tikzpicture}
	\caption{BCJR-once rates $R_{\text{BCJR-once}}$ versus $p_\I=p_\D$ with $p_\S=0$ and $M=1$ using different inner codes.}
	\label{fig:BCJR-Once-Inner}
	 \end{figure}%

Fig.~\ref{fig:BCJR-Once-Inner} compares the BCJR-once rates of some inner codes we will use in this paper along with an inner code designed in \cite{Srinivasavaradhan2021TrellisBMA} for the single sequence scenario. The parameters of the inner codes are given in Table~\ref{tab:Inner code selection}. These plots have been simulated with $10^5$ channel uses and have been smoothed over several iterations. Interestingly, we can observe that it seems that codes that perform well for small insertion and deletion error probabilities perform poor for large error rates and vice versa. While we do not conjecture this to be a general result, this can be explained for our codes at hand as follows. For small error probabilities, the distance spectrum within a single code block, i.e., the set of possible length-$n$ words that can be generated within one trellis section, is the dominant property influencing the performance, as synchronization between blocks is not a problem. However, on the other hand, for large error rates, codes that can retain synchronization between blocks provide the highest BCJR-once rates. In particular, codes for which the codewords in one block have a small $d_{\mathsf{L}}^{\text{min}}$, such as the watermark code, synchronize well, as the structure of a block is less variable over codewords within one block, helping the receiver to know the rough structure of this block.

The above exposition also indicates in which cases the employment of a random offset sequence can improve the performance of a code. In particular, adding a random sequence has three effects on the inner code. First, while the Hamming distance spectrum is invariant to offsets, the Levenshtein distance spectrum can both improve or worsen when adding an offset sequence. Second, a convolutional code, as a block code, is block-cyclic, meaning that shifting an inner codeword by $n$ symbols to the left or right will again be a valid codeword. It is not surprising that such a property is problematic for synchronization and thus a random offset sequence, which destroys the cyclicity, improves the performance at higher error rates. We can identify this behavior for  the TVC-$2$ inner coding scheme, whose performance worsens compared to TVC-$1$ for small error probabilities, as the carefully designed Levenshtein distance spectrum is destroyed when adding a random sequence. However, for large error probabilities, the random sequence  helps as it produces a larger spectrum of possible codewords, increasing the distinguishability between close blocks. We refer the reader to Section \ref{sec:inner:code} for an in-depth discussion of inner code design.

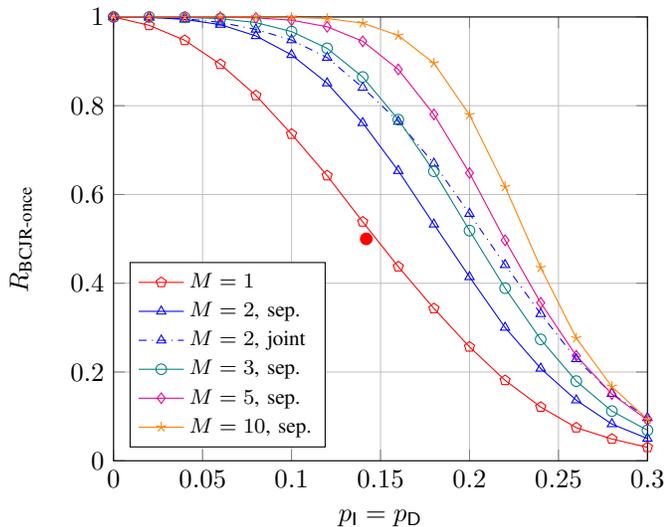
\begin{figure}[t]
 \centering
	\begin{tikzpicture}
		\begin{axis}[
			width = 0.98\columnwidth,
			xmin=0.0,   xmax=0.3,
			ymin=0,	ymax=1,
			xtick distance = 0.05,
			xticklabel style = {/pgf/number format/fixed, /pgf/number format/precision=2},
			grid = both,
			grid style = {line width=.1pt},
			legend pos=south west,
			legend cell align={left},
			legend style={font=\footnotesize},
			xlabel = {$p_\I=p_\D$},
			ylabel style = {name=AIR,yshift=-.0cm},
			ylabel = {$R_{\text{BCJR-once}}$},
			]
			
			\addplot [color=red,mark=pentagon,mark options={solid},each nth point=2] table [col sep=space] {Figures/BCJR-Once/BO-TV4-O-M1.txt};
			\addlegendentry{$M=1$};
			
			\addplot [color=blue,mark=triangle,mark options={solid},each nth point=2] table [col sep=space] {Figures/BCJR-Once/BO-TV4-O-M2.txt};
			\addlegendentry{$M=2$, sep.};
			
			\addplot [color=blue,mark=triangle,mark options={solid},each nth point=2,dashdotted] table [col sep=space] {Figures/BCJR-Once/BO-TV4-O-M2-joint.txt};
			\addlegendentry{$M=2$, joint};
			
			\addplot [color=teal, mark=o,mark options={solid},each nth point=2] table [col sep=space] {Figures/BCJR-Once/BO-TV4-O-M3.txt};
			\addlegendentry{$M=3$, sep.};
			
			\addplot [color=magenta, mark=diamond,mark options={solid},each nth point=2] table [col sep=space] {Figures/BCJR-Once/BO-TV4-O-M5.txt};
			\addlegendentry{$M=5$, sep.};
			
			\addplot [color=orange, mark=star,mark options={solid},each nth point=2] table [col sep=space] {Figures/BCJR-Once/BO-TV4-O-M10.txt};
			\addlegendentry{$M=10$, sep.};
			
			\draw [fill=red,draw=red,thick] (axis cs: 0.142,0.5) circle (2pt);
			
		\end{axis}
	\end{tikzpicture}
	\caption{BCJR-once rates $R_{\text{BCJR-once}}$ versus $p_\I=p_\D$ with $p_\S=0$ using the TVC-$2$ inner coding scheme with different number of transmissions $M$. The red solid circle indicates the iterative decoding threshold of our  concatenated coding scheme with the optimized outer protograph LDPC code from Section~\ref{sec:sim_outer_code_optim} for $M=1$.}
	\label{fig:BCJR-Once-Multiple}
	
 \end{figure}

Fig.~\ref{fig:BCJR-Once-Multiple} shows the BCJR-once rates for the TVC-$2$ inner coding scheme with different number of transmissions $M$. It is observed that multiple transmissions can significantly improve the BCJR-once rates. In particular, even going from one to two sequences, we already see a notable difference in terms of information rates. In addition, as expected, we observe a reduced achievable rate with separate decoding compared to joint decoding. However, this loss can be compensated by decoding more received sequences. For instance, for $p_\I = p_\D = 0.16$, the BCJR-once rate of separate decoding for $M = 3$  is close to that of joint decoding for $M=2$, but with a much lower decoding complexity. In Fig.~\ref{fig:BCJR-Once-Multiple}, we also show the iterative decoding threshold  of our concatenated coding scheme (red solid circle) using the TVC-$2$ inner coding scheme concatenated with an optimized outer   protograph LDPC  code for $M=1$. We observe that (in the asymptotic limit of infinitely-large block length) the proposed coding scheme performs very close to the corresponding AIR. The optimized base matrix  found by computer search along with the corresponding iterative decoding threshold are given in Section~\ref{sec:sim_outer_code_optim}.

\begin{figure}[t]
 \centering
	\begin{tikzpicture}
		\begin{axis}[
			width = 0.98\columnwidth,
			xmin=0.0,   xmax=0.3,
			ymin=0,	ymax=1,
			xtick distance = 0.05,
			xticklabel style = {/pgf/number format/fixed, /pgf/number format/precision=2},
			grid = both,
			grid style = {line width=.1pt},
			legend pos=south west,
			legend cell align={left},
			legend style={font=\footnotesize},
			xlabel = {$p_\I=p_\D$},
			ylabel style = {name=AIR,yshift=-.0cm},
			ylabel = {$R_{\text{BCJR-once}}$},
			]
			
			\addplot [color=blue,mark=square,mark options={solid},each nth point=2] table [col sep=space] {Figures/BCJR-Once/CC2-O-PE0-M1.txt};
			\addlegendentry{$p_\S = 0$};
			
			\addplot [color=red,mark=pentagon, mark options={solid},each nth point=2] table [col sep=space] {Figures/BCJR-Once/CC2-O-PE005-M1.txt};
			\addlegendentry{$p_\S = 0.05$};
			
			\addplot [color=orange,mark=triangle,  mark options={solid},each nth point=2] table [col sep=space] {Figures/BCJR-Once/CC2-O-PE01-M1.txt};
			\addlegendentry{$p_\S = 0.1$};
			
			\addplot [color=blue,mark=square, dashed, mark options={solid},each nth point=2] table [col sep=space] {Figures/BCJR-Once/CC2-O-PE0-M2.txt};

			\addplot [color=red,mark=pentagon, dashed, mark options={solid},each nth point=2] table [col sep=space] {Figures/BCJR-Once/CC2-O-PE005-M2.txt};

			\addplot [color=orange,mark=triangle, dashed,mark options={solid},each nth point=2] table [col sep=space] {Figures/BCJR-Once/CC2-O-PE01-M2.txt};

			\addplot [color=blue,mark=square, dashdotted, mark options={solid},each nth point=1] table [col sep=space] {Figures/BCJR-Once/CC2-O-PE0-M2-joint.txt};

			\addplot [color=red,mark=pentagon, dashdotted, mark options={solid},each nth point=1] table [col sep=space] {Figures/BCJR-Once/CC2-O-PE005-M2-joint.txt};

			\addplot [color=orange,mark=triangle, dashdotted,mark options={solid},each nth point=1] table [col sep=space] {Figures/BCJR-Once/CC2-O-PE01-M2-joint.txt};

			\draw [fill=blue,draw=blue,thick] (axis cs: 0.134,0.5) circle (2pt);
			\draw [fill=red,draw=red,thick] (axis cs: 0.105,0.5) circle (2pt);
			\draw [fill=orange,draw=orange,thick] (axis cs: 0.075,0.5) circle (2pt);
			
		\end{axis}
	\end{tikzpicture}
	\caption{BCJR-once rates $R_{\text{BCJR-once}}$ versus $p_\I=p_\D$ with $p_\S=0,0.05,0.1$ using the CC-$2$ inner coding scheme with $M=1$ and $M=2$. Solid lines are for $\numS=1$, dashed lines are for $\numS=2$ and separate decoding, and dash dotted lines are for $\numS=2$ and joint decoding. The solid circles indicate the iterative decoding thresholds of our  concatenated coding scheme with the optimized outer protograph LDPC codes from Section~\ref{sec:FER:substitutions} for $M=1$.}
	\label{fig:BCJR-Once-Multiple-Ps}
	
 \end{figure}
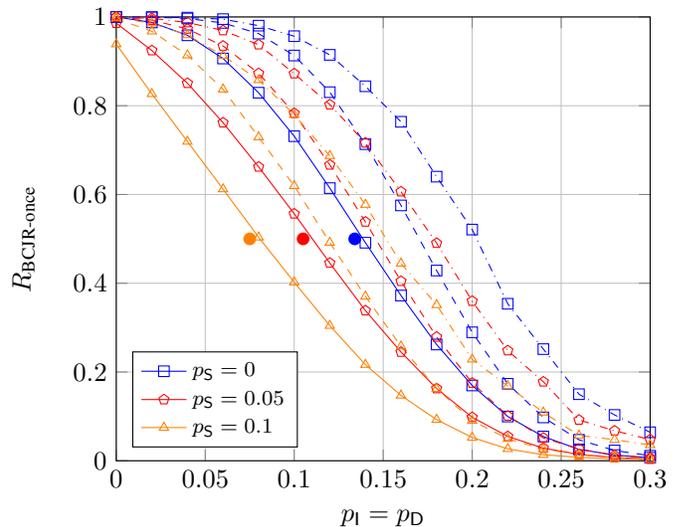

In Fig.~\ref{fig:BCJR-Once-Multiple-Ps}, we show BCJR-once rates with an inner convolutional code (CC-$2$) for both single and double sequence transmission (separate and joint decoding) when $p_{\mathsf{S}} = 0, 0.05, 0.1$. As can be observed from the figure, the achievable rate loss of separate decoding with increasing $p_{\mathsf{S}}$ is about the same for $M=1$ and $M=2$. Moreover,  the gap between separate and joint decoding (for $M=2$) stays approximately the same for different $p_{\mathsf{S}}$, which shows that the proposed separate decoding approach is robust to substitution errors. In the figure we also show the iterative decoding thresholds of our concatenated coding scheme (solid circles)  using the CC-$2$ inner coding scheme with optimized   outer protograph LDPC  codes, optimized individually for each $p_{\mathsf{S}}$, for $M=1$ (see Section~\ref{sec:FER:substitutions} for the optimized base matrices).

\subsection{Mutual Information Rate}
A  method to compute the mutual information for a given coding scheme was introduced in  \cite{pfister_achievable_2001,arnold_simulation-based_2006}, where the mutual information between an input process $\w = (w_1, w_2, \dots)$ and an output process $\y = (y_1, y_2, \dots)$, $I(\w;\y)$, is computed via trellis-based simulations. Given that a source/channel decoding trellis exists for a coding scheme, the mutual information point
\begin{align} \label{eq:MIrate}
I(w;y) \triangleq \lim_{\leno \to \infty} \frac{1}{\leno+m} I(\w;\y),
\end{align}
which is an AIR, can be computed using the forward recursion of the BCJR algorithm on the given trellis. Since
\begin{align*} 
	I(\w; \y) = H(\w) + H(\y) - H(\w, \y)
\end{align*}
and $\log p(\w)$, $\log p(\y)$, and $\log p(\w,\y)$ converge with probability 1 to $ H(\w)$,  $H(\y)$, and $H(\w, \y)$, respectively, for long sequences, $I(w;y)$ can be estimated by

\begin{align*}
	\hat{I}(w; y) = & -\frac{1}{\leno + m}\log p(\w) -\frac{1}{\leno + m}\log p(\y)\\ &+\frac{1}{\leno + m}\log p(\w,\y)
\end{align*}
when $\leno$ is very large.

This approach can be used for our coding schemes as well. For the case of single sequence transmission, given an input sequence $\w$ to the inner code and a corresponding output sequence $\y$ from the channel, $\log p(\y)$, $\log p(\w,\y)$, and $\log p(\w)$ can be computed as follows. First, compute $\log p(\y)$ from
\begin{align*}
	p(\y) = \sum_{\sigma} p \bigl(\y_{1}^{(\leno+m)n+d}, \sigma \bigr) \overset{(a)}{=} \sum_{\sigma} \alpha_{\leno + m}(\sigma),
\end{align*}
where $(a)$ follows since $\alpha_{i}(\sigma) = p\bigl(\y_{1}^{in+d}, \sigma \bigr)$. Hence,  it can be computed using the recursion in \eqref{eq:alpha_rec}. Second,  $\log p(\w)$ and $\log p(\w,\y)$ can be computed using a recursion in a similar manner. In particular, the recursion for computing $p(\w,\y)$ is
\begin{align*}
	\alpha_{i}^{(\mathsf{\w,\y})}(\sigma) = \sum_{\hat{\sigma}} \alpha_{i-1}^{(\mathsf{\w,\y})}(\hat{\sigma}) \gamma_i(\hat{\sigma},\sigma),
\end{align*}
where the summation is over all states $\hat{\sigma}$  with an outgoing edge to $\sigma$ labeled with the input sequence symbol $w_i$ at time $i$. In other words, the recursion for $p(\w,\y)$ does not marginalize the input sequence $\w$ like in $p(\y)$. Then, 
\begin{align*}
	p(\w, \y) = \sum_{\sigma} \alpha^{(\mathsf{w,y})}_{\leno + m}(\sigma).
\end{align*}
The recursion for computing $p(\w)$ is 
\begin{align*}
	\alpha_{i}^{(\mathsf{\w})}(\sigma) = \sum_{\hat{\sigma}} \alpha_{i-1}^{(\mathsf{\w})}(\hat{\sigma}) p(w_i, \sigma |\hat{\sigma}),
\end{align*}
where again the summation is over all states $\hat{\sigma}$  with an outgoing edge to $\sigma$ labeled with the input sequence symbol $w_i$ at time $i$. 
Then, 
\begin{align*}
	p(\w)=
	\sum_{\sigma} \alpha^{(\mathsf{w})}_{\leno + m}(\sigma).
\end{align*}
However, since we consider an input sequence of independent and uniformly distributed symbols,  $H(\w)$ is equal to $(\leno + m)\log \outq$, and hence we do not need to run the recursion.

\begin{figure}[t]
    \centering
    \begin{tikzpicture}
	\begin{axis}[
		width = 0.98\columnwidth,
		xmin=0.0,   xmax=0.3,
		ymin=0,	ymax=1,
		xtick distance = 0.05,
		xticklabel style = {/pgf/number format/fixed, /pgf/number format/precision=2},
		grid = both,
		grid style = {line width=.1pt},
		legend pos=south west,
		legend cell align={left},
		legend style={font=\footnotesize},
		xlabel = {$p_\I=p_\D$},
		ylabel style = {name=AIR,yshift=-.0cm},
		ylabel = {$R_{\text{MI}}$},
		]
		
		\addplot [color=black, mark = diamond, each nth point=2] table [col sep=space] {Figures/Arnold-Loeliger/AIR_DNAChannel_M1_AL_FullLattice.txt};
		\addlegendentry{Uncoded};
		
		\addplot [color=blue, mark = square,each nth point=2] table [col sep=space] {Figures/Arnold-Loeliger/AIR_CC_DNAChannel_M1_AL.txt};
		\addlegendentry{CC-$1$/CC-$2$};
		
		\addplot [color=red, mark = star, each nth point=2] table [col sep=tab] {Figures/Arnold-Loeliger/AIR_WM_DNAChannel_M1_AL.txt};
		\addlegendentry{WM};
		
		\addplot [color=cyan, mark = triangle, mark options={solid},each nth point=2] table [col sep=tab] {Figures/Arnold-Loeliger/AIR_TVC_DNAChannel_M1_AL.txt};
		\addlegendentry{TVC-$1$};
		
		\addplot [color=orange,mark = o, each nth point=2] table [col sep=tab] {Figures/Arnold-Loeliger/AIR_TVC+RS_DNAChannel_M1_AL.txt};
		\addlegendentry{TVC-$2$};
		
	\end{axis}
\end{tikzpicture}
	
    \caption{Mutual information rates $R_{\text{MI}}$ versus $p_\I=p_\D$ with $p_\S=0$ for single sequence transmission for several inner coding schemes.}
    \label{fig:A&L AIRs}
\end{figure}
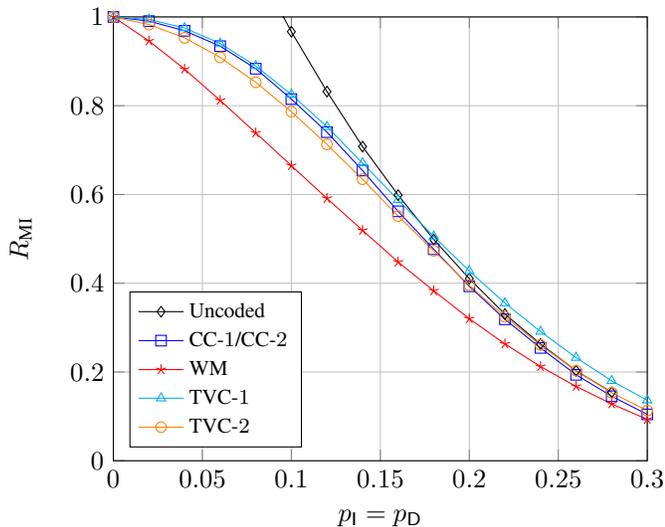

Similarly, this approach can be used to compute the analog mutual information point of \eqref{eq:MIrate}  for the case of multiple received sequences. Given an input sequence  $\w$ transmitted over $M$ identical and independent IDS channels, we will receive the $M$ output sequences $\y_1,\dots,\y_{\numS}$. Then, the analog mutual information point of \eqref{eq:MIrate}, denoted by $I(w;y_1,\dots,y_M)$, which is an AIR, can be estimated by
\begin{align*}
	\hat{I}(w;y_{1},\dots,y_{\numS}) = &-\frac{1}{\leno + m}\log p(\w)\\ &-\frac{1}{\leno + m}\log p(\y_{1},\dots,\y_{\numS}) \\
	&+\frac{1}{\leno + m}\log p(\w,\y_{1},\dots,\y_{\numS})
\end{align*}
when $\leno$ is very large. By considering the joint sequences  $(\y_1)_{1}^{in+d_1},\dots,(\y_M)_{1}^{in+d_M}$, $\log p(\w)$, $\log p(\y_{1},\dots,\y_{\numS})$, and $\log p(\w,\y_{1},\dots,\y_{\numS})$ can be computed in an analog way as for the single sequence case.

\begin{figure}[t]
    \centering
    \begin{tikzpicture}
	\begin{axis}[
		width = 0.98\columnwidth,
		xmin=0.0,   xmax=0.3,
		ymin=0,	ymax=1,
		xtick distance = 0.05,
		xticklabel style = {/pgf/number format/fixed, /pgf/number format/precision=2},
		grid = both,
		grid style = {line width=.1pt},
		legend pos=south west,
		legend cell align={left},
		legend style={font=\footnotesize},
		xlabel = {$p_\I=p_\D$},
		ylabel style = {name=AIR,yshift=-.0cm},
		ylabel = {$R_{\text{MI}}$},
		]
		
		\addplot [color=black, mark = diamond, each nth point=2] table [col sep=space] {Figures/Arnold-Loeliger/AIR_DNAChannel_M2_AL.txt};
		\addlegendentry{Uncoded};
		
		\addplot [color=blue, mark = square,each nth point=2] table [col sep=space] {Figures/Arnold-Loeliger/AIR_CC_DNAChannel_M2_AL.txt};
		\addlegendentry{CC-$1$/CC-$2$};
		
		\addplot [color=red, mark = star, each nth point=2] table [col sep=space] {Figures/Arnold-Loeliger/AIR_WM_DNAChannel_M2_AL.txt};
		\addlegendentry{WM};
		
		\addplot [color=cyan, mark = triangle, mark options={solid},each nth point=2] table [col sep=space] {Figures/Arnold-Loeliger/AIR_TVC_DNAChannel_M2_AL.txt};
		\addlegendentry{TVC-$1$};
		
				\addplot [color=orange, mark = o, mark options={solid},each nth point=2] table [col sep=space] {Figures/Arnold-Loeliger/AIR_TVC+RS_DNAChannel_M2_AL.txt};
		\addlegendentry{TVC-$2$};
		
	\end{axis}
\end{tikzpicture}
	
    \caption{Mutual information rates $R_{\text{MI}}$ versus $p_\I=p_\D$ with $p_\S=0$  for multiple sequence transmission with $\numS = 2$ for several inner coding schemes.}
    \label{fig:A&L AIRs M=2}
\end{figure}

Fig.~\ref{fig:A&L AIRs} shows the mutual information rate $R_{\text{MI}} \approx \hat{I}(w;y)$ for the different inner codes that we consider in this paper (see Table~\ref{tab:Inner code selection}). These plots have been simulated using $10^6$ channel uses. It is evident from the curves that the TVC-$1$ scheme performs the best for high insertion and deletion probabilities. The TVC-$1$ scheme even outperforms the uncoded scheme, which might seem surprising at first glance; however, this  can happen when memory is introduced to the source/channel, which is the case for TVC-$1$. The mutual information rate gives us insights on the expected performance of the tabulated inner codes concatenated with an outer code tailored toward them, where extrinsic information has been passed between the inner and outer decoders. In other words, the achievable rates $R_{\text{MI}}$ give insights on the inner-outer iterative decoding performance of our coding scheme. This is the reason why we see a difference between mutual information and BCJR-once rates ($R_{\text{MI}} \geq R_{\text{BCJR-once}}$), where we also observe a change in which inner code achieves the better rate. This is expected since the effects of synchronization loss can be mitigated by performing inner-outer decoding. We observe and confirm this disposition in our results section. Finally, the achievable rate $R_{\text{MI}} \approx \hat{I}(w;y_1,\dots,y_M)$ for the case of multiple sequence transmission with $M = 2$ is shown in Fig.~\ref{fig:A&L AIRs M=2}.

It should be noted that the rate $R_{\text{MI}}$ for the uncoded case in Figs.~\ref{fig:A&L AIRs} and \ref{fig:A&L AIRs M=2} is computed using the lattice implementation introduced in Section~\ref{sec:innerdec-one}. Since in our decoding algorithm we have to limit the maximum and minimum drift values, we end up with inaccurate computations of $\log p(\y)$ and $\log p(\w,\y)$. By using a lattice implementation, we can exactly compute these quantities (details omitted for brevity). The discrepancy between the two methods is most visible for the uncoded case, whereas it is negligible for the other inner codes.  

%

%
\section{Simulation Results} \label{sec:simresults}
In this section, we provide  frame error rate (FER) performance results for several set-ups of our coding schemes. First, we show a comparison of the different inner coding schemes introduced in the earlier sections. In particular, we show that our results are in accordance with the BCJR-once and mutual information rates. Second, we simulate the FER performance of our designed nonbinary polar and LDPC codes with the TVC-$2$ inner coding scheme. Third, we simulate our coding schemes under short sequence transmission conditions. Next, we combine the best inner and outer code selection and simulate for the case of multiple received sequences. In addition, we compare the performance of our coding schemes in the multiple sequence transmission scenario with an existing  MSA algorithm.  Finally, to show  robustness to substitution errors, we present FER  results for two different values of $p_\S>0$ using the CC-$2$ inner coding scheme concatenated with a designed  nonbinary LDPC code.

Our simulations are done over an alphabet of size $q=4$, which equates to the four bases  $\{ \mathsf{A}, \mathsf{C}, \mathsf{G}, \mathsf{T} \}$ of the DNA and we set $p_\S = 0$ (except for Fig.~\ref{fig:FER_subsErrors}) and $p_\I=p_\D$. An outer LDPC code is decoded via belief propagation with a maximum number of $100$ iterations. When applicable, the maximum number of iterations between the inner and outer decoders is set to $100$. The polar code is decoded via CRC-aided SCL decoding. Regarding the inner codes, we set $I_\mathrm{max} = 2$ and the limit of the drift random variable in decoding is set dynamically as follows. We set the drift limit to five times the standard deviation of the final drift at position $\len$, i.e., $d_{\max} = -d_\mathrm{min}  = 5 \sqrt{\vphantom{A}\smash{N \frac{p_\D}{1-p_\D}}}$. However, if the received sequence has a drift outside this limit, we increase the limit to be ten times the standard deviation. This can be motivated by the fact that in DNA storage, the length of each strand is known, so the drift limit in decoding can be set accordingly. Moreover, for multiple transmissions, we use $\numS = 2$, $3$, $5$, and $10$. For short sequence transmission, our overall code length is $N = 128$ DNA symbols, while we use $N = 960$ DNA symbols otherwise. We picked the short and long sequence  lengths to be within the range of the corresponding DNA sequencing technologies. Illumina sequencing can produce sequence lengths ranging between $100-300$ DNA symbols, while Oxford nanopore sequencing produces sequences of lengths $1000-2000$ DNA symbols. All FER results are with an overall code rate of $R = \nicefrac{1}{2}$ (in bits per DNA symbol), with an outer code rate of $\rateo = \nicefrac{1}{2}$ and an inner code rate of $\ratei = 1$.

\subsection{Inner Code Optimization/Comparison} \label{sec:inner:code:optimization}
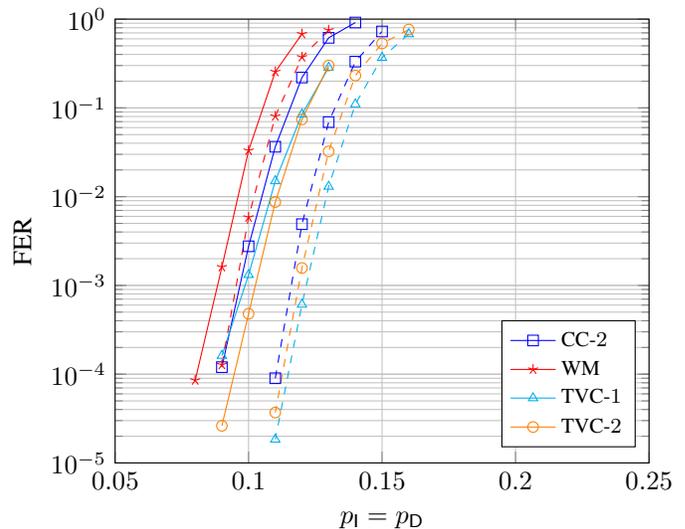
\begin{figure}[t]
    \centering
    \begin{tikzpicture}[scale=1.0]
\begin{semilogyaxis}[
width = 0.98\columnwidth,
xmin=0.05,   xmax=0.25,
ymin=1e-5,	ymax=1,
xticklabel style = {/pgf/number format/fixed, /pgf/number format/precision=6},
xtick={0.05,0.1,0.15,0.2, 0.25},
ymode=log,
grid = both,
grid style = {line width=.1pt},
legend cell align={left},
legend style={font=\footnotesize,at={(axis cs: 0.245,1.5E-5)},anchor=south east},
xlabel = {$p_\I=p_\D$},
ylabel = {FER},
cycle list name=color list
]

\addplot+ [color=blue, mark=square] table [col sep=comma, x=p, y=FER] {Figures/FER_innerWiMax/BCJR_CC_WIMAXlike_q16_n240_k120_M1.csv};
\addlegendentry{CC-$2$};

\addplot+ [color=red, mark=star] table [col sep=space, x=pins, y=FER] {Figures/FER_innerWiMax/WMnw8,kw4+Wimax-like_NB-LDPCn240,k120_dv6_dc7_GF16.txt};
\addlegendentry{WM};

\addplot+ [color=cyan, mark=triangle] table [col sep=space, x=pins, y=FER] {Figures/FER_innerWiMax/RandomTVCnw8,kw4+Wimax-like_NB-LDPCn240,k120_dv6_dc7_GF16.txt};
\addlegendentry{TVC-$1$};

\addplot+ [color=orange, mark=o] table [col sep=space, x=pins, y=FER] {Figures/FER_innerWiMax/TVC+RSnw8,kw4+Wimax-like_NB-LDPCn240,k120_dv6_dc7_GF16.txt};
\addlegendentry{TVC-$2$};

\addplot+ [color=red, mark=star, dashed] table [col sep=space, x=pins, y=FER] {Figures/FER_innerWiMax/WMnw8,kw4+Wimax-like_NB-LDPCn240,k120_dv6_dc7_GF16_it.txt};

\addplot+ [color=cyan, mark=triangle, dashed,mark options={solid}] table [col sep=space, x=pins, y=FER] {Figures/FER_innerWiMax/RandomTVCnw8,kw4+Wimax-like_NB-LDPCn240,k120_dv6_dc7_GF16_it.txt};

\addplot+ [color=blue, mark=square, dashed, mark options={solid}] table [col sep=comma, x=p, y=FER] {Figures/FER_innerWiMax/BCJR_CC_WIMAXlike_q16_n240_k120_M1_turbo100.csv};

\addplot+ [color=orange, mark=o, dashed, mark options={solid}] table [col sep=space, x=pins, y=FER] {Figures/FER_innerWiMax/TVC+RSnw8,kw4+Wimax-like_NB-LDPCn240,k120_dv6_dc7_GF16_it.txt};

\end{semilogyaxis}%
\end{tikzpicture}%
    \caption{FER performance vs. $p_\I=p_\D$   on different inner codes concatenated with a $[240,120]_{2^4}$ WiMax-like outer LDPC  code with overall block length of $\len = 960$ DNA symbols and with $p_\S = 0$. Dashed lines are with inner-outer iterations while solid lines are without.}
    \label{fig:Inner-code comp}
\end{figure}

Comparing the different inner codes in our coding schemes can be done in several ways. We have already presented one way of comparison by providing the BCJR-once and mutual information rates. However, these rates correspond to the asymptotics of these codes. To present a more clear comparison, we plot in Fig.~\ref{fig:Inner-code comp} the FER performance of the different inner coding schemes concatenated with a $[240, 120]_{2^4}$ WiMax-like nonbinary outer LDPC  code. Confirming the BCJR-once rate results in Fig.~\ref{fig:BCJR-Once-Inner}, for a coding rate of $\nicefrac{1}{2}$, we observe that the TVC-$2$ inner coding scheme performs the best with no iterations between the inner and outer decoders. Furthermore, when considering iterative inner-outer decoding, the TVC-$1$ scheme performs the best. This in turn confirms the mutual information rate results in Fig.~\ref{fig:A&L AIRs}. It is evident that our designed inner code improves the performance of the overall concatenated coding scheme.

\begin{table} 
	\setlength{\tabcolsep}{1.7pt}
	\begin{center}
		\caption{Iterative Decoding Thresholds  of  LDPC Codes With $p_{\S}=0$}
		\vspace{-2ex}
		{\renewcommand{\arraystretch}{1.1}
			\begin{tabular}{ccc} \specialrule{1.2pt}{0pt}{0pt}
				Code & $p_{\mathsf{th}}$ & Rate \\ \specialrule{.8pt}{0pt}{0pt}
				Designed & 0.142 & $\nicefrac{1}{2}$ \\
				WiMax-like  & 0.139 & $\nicefrac{1}{2}$ \\
				\specialrule{1.2pt}{0pt}{4pt}
			\end{tabular}\label{tab:Thresholds of LDPC codes}}
	\end{center}
	\vspace{-.4cm}
\end{table}

\subsection{Outer Code Optimization} \label{sec:sim_outer_code_optim}
In the following, we show FER performance results for the TVC-$2$ inner coding scheme  concatenated with our designed outer codes. The optimization of the outer code was presented in Sections~\ref{sec:outerLDPC} and \ref{sec:outerPolar}, where we  describe how to design, respectively,  an outer LDPC or polar code tailored to an inner coding scheme combined with the IDS channel. The results are shown in Fig.~\ref{fig:outer-code comp}. Furthermore, Table~\ref{tab:Thresholds of LDPC codes} shows the iterative decoding threshold $p_{\mathsf{th}}$ of the WiMax-like LDPC code and our designed protograph LDPC code. The base matrix corresponding to our designed protograph LDPC code is 
\begin{align} \label{eq:B}
 \boldsymbol{B} = \begin{pmatrix}
   1 & 2 & 1 & 1 \\
   1 & 1 & 2 & 1 \\
    \end{pmatrix}.
\end{align}
It should  be noted that we simulate the ensemble average of the lifted protograph (see Section~\ref{sec:outerLDPC}) by assigning new random weights for the edges of the Tanner  graph for every new block transmission. Also, the LDPC code has been optimized for the case of no iterations between the  inner and outer decoders and for the TVC-$2$ inner coding scheme. Evidently, we succeed in improving the performance of our coding schemes with both of the designed outer codes; the protograph LDPC code performing the best. The designed protograph LDPC code is a $[240, 120]_{2^4}$ code with girth $10$, which gives a  block length of $N = 960$ DNA symbols, while the designed polar code is a $[256, 128]_{2^4}$ code, resulting in $N = 1024$ DNA symbols. Again, as for the LDPC code, the polar code has been optimized for the TVC-$2$ inner coding scheme and with no iterations between the inner and outer decoders. These results validate our optimization and design techniques as we have managed to improve the performance compared to the standard case. For a better visual representation of how well the optimization performs, we plotted in Fig.~\ref{fig:BCJR-Once-Multiple} (the red solid circle) the iterative decoding threshold for our concatenated coding scheme (TVC-$2$ inner coding scheme and optimized outer LDPC code constructed from the base matrix $\boldsymbol{B}$ in \eqref{eq:B}) for $M=1$. The gap to the AIR curve is very small, which validates our optimization. On a side note, optimizing the inner code is of greater importance as the inner code is responsible for maintaining synchronization, which is the main obstacle faced when dealing with IDS channels.

\begin{figure}[t]
    \centering
    \begin{tikzpicture}[scale=1.0]
\begin{semilogyaxis}[
width = 0.98\columnwidth,
xmin=0.05,   xmax=0.25,
ymin=1e-5,	ymax=1,
xticklabel style = {/pgf/number format/fixed, /pgf/number format/precision=6},
xtick={0.05,0.1,0.15,0.2, 0.25},
ymode=log,
grid = both,
grid style = {line width=.1pt},
legend cell align={left},
legend style={font=\footnotesize,at={(axis cs: 0.245,1.5E-5)},anchor=south east},
xlabel = {$p_\I=p_\D$},
ylabel = {FER},
cycle list name=color list
]

\addplot+ [color=red, mark=star] table [col sep=space, x=pins, y=FER] {Figures/FER_innerWiMax/TVC+RSnw8,kw4+Wimax-like_NB-LDPCn240,k120_dv6_dc7_GF16.txt};
\addlegendentry{WiMax-like LDPC};

\addplot+ [color=blue, mark=square, mark options={solid}] table [col sep=comma, x=p, y=FER] {Figures/polar_long/bcjr_TV_polar_a1_k128_n256_q16_l32_c8_ker-61.csv};
\addlegendentry{Opt. polar};

\addplot+ [color=orange, mark=o] table [col sep=space, x=pins, y=FER] {Figures/FER_innerWiMax/TVC+RSnw8,kw4+opt-LDPCn240,k120_protograph_dv3_dc5_GF16.txt};
\addlegendentry{Opt. LDPC};

\end{semilogyaxis}%
\end{tikzpicture}%
    \caption{FER performance vs. $p_\I=p_\D$  on different outer codes concatenated with the TVC-$2$ inner coding scheme, with $p_\S= 0$ and with   no iterations between the inner and outer decoders. The polar code has parameters $\leno = 256$, $\rateo=\nicefrac{1}{2}$, $\outq=16$, $\nicefrac{\alpha}{\beta} = 6$, list size $32$, and $\crclen = 8$.}
    \label{fig:outer-code comp}
\end{figure}
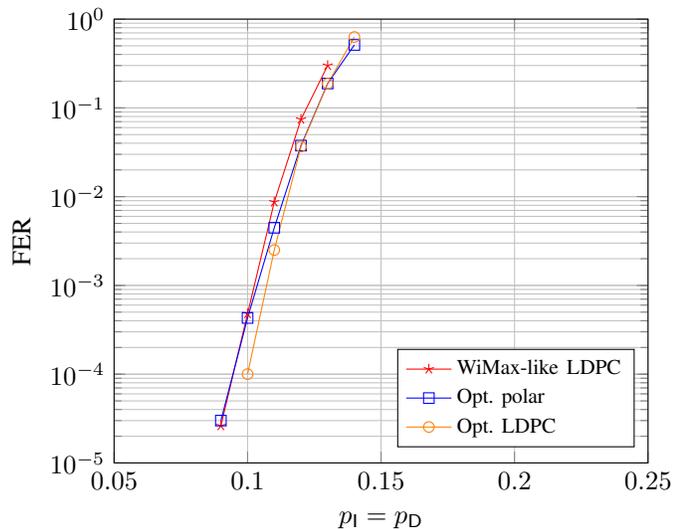

\begin{figure}[t]
     \centering
     \begin{tikzpicture}[scale=1.0]
\begin{semilogyaxis}[
width = 0.98\columnwidth,
xmin=0.02,   xmax=0.14,
ymin=1e-5,	ymax=1,
xticklabel style = {/pgf/number format/fixed, /pgf/number format/precision=6},
ymode=log,
grid = both,
grid style = {line width=.1pt},
legend cell align={left},
legend style={font=\footnotesize,at={(axis cs: 0.135,1.5E-5)},anchor=south east},
xlabel = {$p_\I=p_\D$},
ylabel = {FER},
cycle list name=color list
]

	\addplot+ [color=blue, mark=square] table [col sep=comma, x=p, y=FER] {Figures/Short/BCJR_CC_polar_a1_k32_n64_q4_l32_c8_ker-31_M1_C.csv};
\addlegendentry{CC-$1$};

\addplot+ [color=red, mark=star] table [col sep=comma, x=p, y=FER] {Figures/Short/BCJR_WM_polar_a1_k32_n64_q4_l32_c8_ker-31_M1_C.csv};
\addlegendentry{WM};

\addplot+ [color=cyan, mark=triangle] table [col sep=comma, x=p, y=FER] {Figures/Short/BCJR_TVRS_NoOffset_polar_a1_k32_n64_q4_l32_c8_ker-31_M1_C.csv};
\addlegendentry{TVC-$1$};

\addplot+ [color=cyan, dashed, mark=triangle, mark options={solid}] table [col sep=comma, x=p, y=FER] {Figures/Short/LDPC/BCJR_TVRS_NoOffset_EirikOPTldpc_n64k32q4_innerTVRS_NoOffset_C.csv};

\addplot+ [color=cyan, dashed, mark=triangle, mark options={solid}] table [col sep=comma, x=p, y=FER] {Figures/Short/LDPC/BCJR_TVRS_NoOffset_EirikOPTldpc_n64k32q4_innerTVRS_NoOffset_C.csv};

\addplot+ [color=blue, mark=square, dashed, mark options={solid}] table [col sep=comma, x=p, y=FER] {Figures/Short/LDPC/BCJR_CC_EirikOPTldpc_n64k32q4_innerCC_WithOffset_C.csv};

\addplot+ [color=red, dashed, mark=star, mark options={solid}] table [col sep=comma, x=p, y=FER] {Figures/Short/LDPC/LDPCopt_WM.csv};

\end{semilogyaxis}%
\end{tikzpicture}%
     \caption{FER performance vs. $p_\I=p_\D$ on short sequences of $\len = 128$ DNA symbols for different inner coding schemes,  with $p_\S = 0$ and with no iterations between the inner and outer decoders. Solid lines are for an outer polar code with parameters $\leno = 64$, $\rateo=\nicefrac{1}{2}, \outq=4$, $\nicefrac{\alpha}{\beta} = 3$, list size $32$, and $\crclen = 8$, while dashed lines are for an outer $[64,32]_{2^2}$ optimized LDPC code.}
     \label{fig:polar-short-single}
 \end{figure}
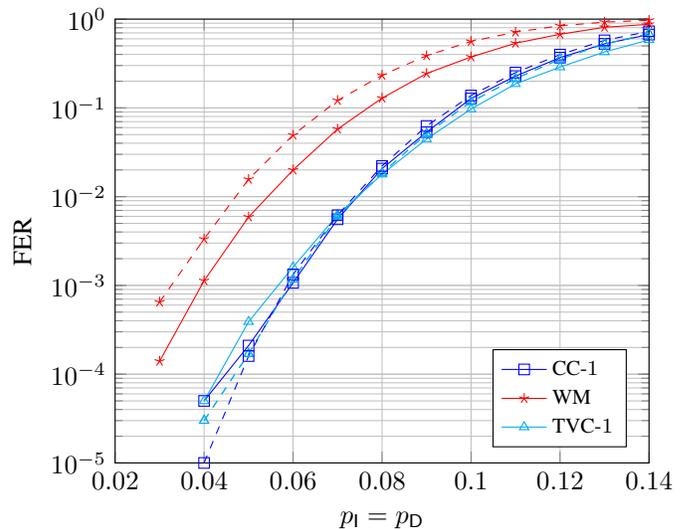

\subsection{Short Sequence Transmission}

The FER performance for the short block length regime is presented in Fig.~\ref{fig:polar-short-single}. The comparison is done on different inner codes using a $[64,32]_{2^2}$ outer polar code or a $[64,32]_{2^2}$ LDPC code that both have been optimized separately for each inner coding scheme. In fact, the  protograph in \eqref{eq:B} is optimal for  all four WM, CC-$1$, TVC-$1$, and TVC-$2$ inner coding schemes, which could be attributed to the rather small search space. The corresponding designed nonbinary LDPC code has girth $8$. Similar to the long sequence case, the watermark code is performing worse than other inner code choices, which has been as well predicted by the AIR results for our chosen rate. Surprisingly, other combinations of inner and outer codes perform very similar in the short sequence case. Due to clarity of presentation, we have excluded in the plot the results for the TVC-$2$ inner coding scheme, which performed the best for long sequences. However, the performance of this coding scheme is slightly worse than the CC-$1$ and TVC-$1$ results. Moreover, the TVC-$1$ inner coding scheme performs better for higher insertion and deletion probabilities than all inner codes with a random offset sequence. The difference of the results compared to the AIR and long sequence results may stem from the fact that for short sequences the synchronization process is done over a shorter trellis. Therefore, the knowledge of the fixed start and end point has more influence on the APPs in the middle of the sequence, which makes synchronization without a random offset sequence also more feasible for higher insertion and deletion probabilities. However, the CC-$1$ inner coding scheme seems to perform best for decreasing failure probabilities for short sequences. Considering the outer code choice, for the TVC-$1$ and CC-$1$ inner coding schemes, the LDPC code and the polar code yield similar performance, with the polar code slightly outperforming  the optimized nonbinary LDPC code for high deletion and insertion probabilities and vice versa for low error probabilities. However, the polar code outperforms the LDPC code by a more significant gap for the WM inner coding scheme. This may be explained by the limited search space for the LDPC protograph; a higher dimension protograph might allow to close the gap.

\begin{figure}[t]
    \centering
    \begin{tikzpicture}[scale=1.0]
\begin{semilogyaxis}[
width = 0.98\columnwidth,
xmin=0.05,   xmax=0.25,
ymin=1e-5,	ymax=1,
xticklabel style = {/pgf/number format/fixed, /pgf/number format/precision=6},
ymode=log,
grid = both,
grid style = {line width=.1pt},
legend cell align={left},
legend style={font=\footnotesize,at={(axis cs: 0.055,1.5E-5)},anchor=south west},
xlabel = {$p_\I=p_\D$},
ylabel = {FER},
cycle list name=color list
]

\addplot+ [color=red, mark=pentagon] table [col sep=space, x=pins, y=FER] {Figures/Multiple_sequences/TVC+RSnw8,kw4+opt-LDPCn240,k120_protograph_dv3_dc5_GF16.txt};
\addlegendentry{$M = 1$};

\addplot+ [color=blue, mark=triangle] table [col sep=space, x=pins, y=FER] {Figures/Multiple_sequences/TVC+RSnw8,kw4+opt-LDPCn240,k120_protograph_dv3_dc5_GF16_M2.txt};
\addlegendentry{$M = 2$, sep.};

\addplot+ [color=blue, mark=triangle, mark options={solid}, dashdotted] table [col sep=space, x=pins, y=FER] {Figures/Multiple_sequences/TVC+RSnw8,kw4+opt-LDPCn240,k120_protograph_dv3_dc5_GF16_M2_joint.txt};
\addlegendentry{$M = 2$, joint};

\addplot+ [color=teal, mark=o] table [col sep=space, x=pins, y=FER] {Figures/Multiple_sequences/TVC+RSnw8,kw4+opt-LDPCn240,k120_protograph_dv3_dc5_GF16_M3.txt};
\addlegendentry{$M = 3$, sep.};

\addplot+ [color=magenta, mark=diamond] table [col sep=space, x=pins, y=FER] {Figures/Multiple_sequences/TVC+RSnw8,kw4+opt-LDPCn240,k120_protograph_dv3_dc5_GF16_M5.txt};
\addlegendentry{$M = 5$, sep.};

\addplot+ [color=orange, mark=star] table [col sep=space, x=pins, y=FER] {Figures/Multiple_sequences/TVC+RSnw8,kw4+opt-LDPCn240,k120_protograph_dv3_dc5_GF16_M10.txt};
\addlegendentry{$M = 10$, sep.};

\end{semilogyaxis}%
\end{tikzpicture}%
    \caption{FER performance vs. $p_\I=p_\D$ for the TVC-$2$ inner coding scheme concatenated with our optimized $[240,120]_{2^4}$ outer LDPC  code with block length $N = 960$ DNA symbols for multiple sequences, with $p_\S = 0$ and with no iterations between the inner and outer decoders. Solid lines represent separate decoding while dash dotted are for joint decoding.}
    \label{fig:opt_Inner+Outer_Mseq}
\end{figure}
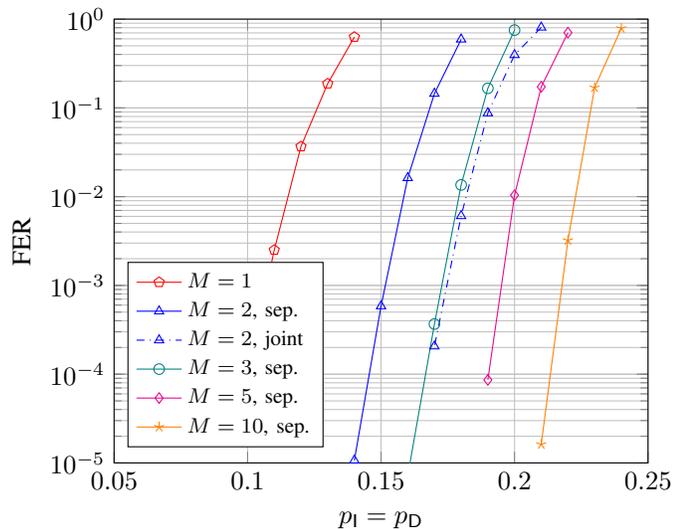

\subsection{Multiple Sequence Transmission} \label{sec:mul:seq:trans}
Next, we present FER results for the case of multiple sequence transmission with both long and short sequences. For transmission block length $N = 960$ DNA symbols, we have shown that our designed protograph LDPC code combined with the TVC-$2$ inner coding scheme performs the best when we do not iterate between the inner and outer decoders. Fig.~\ref{fig:opt_Inner+Outer_Mseq} shows the FER results for this scheme with multiple sequences, where we observe a significant gain in performance when increasing the number of sequences. There is also a notable  gain of  joint decoding compared to  separate decoding as illustrated for $M=2$ sequences. Moreover, separate decoding with $M = 3$ performs similar to optimal joint decoding for $M = 2$, but with much  lower complexity. This confirms the earlier observation with the BCJR-once rates (see Fig.~\ref{fig:BCJR-Once-Multiple}), and shows the viability of our proposed sub-optimal decoding algorithm. For short sequence transmission we perform our analysis with an outer polar code using the CC-$1$ inner coding scheme. In the following  we include a comparison of that coding scheme  with an existing MSA method from the literature for the multiple sequence scenario.\footnote{Note that we have \emph{not} optimized neither the LDPC code nor the polar code for the multiple sequence scenario, but rather use the optimized codes for the case of a single sequence transmission.}

\begin{figure}[t]
    \centering
    \begin{tikzpicture}[scale=1.0]
\begin{semilogyaxis}[
width = 0.98\columnwidth,
xmin=0.05,   xmax=0.25,
ymin=1e-5,	ymax=1,
xticklabel style = {/pgf/number format/fixed, /pgf/number format/precision=6},
ymode=log,
grid = both,
grid style = {line width=.1pt},
legend cell align={left},
legend style={font=\footnotesize,at={(axis cs: 0.245,1.5E-5)},anchor=south east, legend columns=1},
xlabel = {$p_\I=p_\D$},
ylabel = {FER},
cycle list name=color list
]

\addplot+ [color=red,mark=pentagon] table [col sep=comma, x=p, y=FER] {Figures/Short_MultiSeq/bcjr_CC_polar_a1_k32_n64_q4_l32_c8_ker-31.csv};
\addlegendentry{ $M=1$};

\addplot+ [color=blue,mark=triangle] table [col sep=comma, x=p, y=FER] {Figures/Short_MultiSeq/bcjr_CC_polar_a1_k32_n64_q4_l32_c8_ker-31_M2.csv};
\addlegendentry{ $M=2$, sep.};

\addplot+ [color=blue,mark=triangle, dashdotted, mark options={solid}] table [col sep=comma, x=p, y=FER] {Figures/Short_MultiSeq/BCJR_CC_polar_a1_k32_n64_q4_l32_c8_ker-31_M2_Joint.csv};
\addlegendentry{$M=2$, joint};

\addplot+ [color=teal, mark=o] table [col sep=comma, x=p, y=FER] {Figures/Short_MultiSeq/bcjr_CC_polar_a1_k32_n64_q4_l32_c8_ker-31_M3.csv};
\addlegendentry{ $M=3$, sep.};

\addplot+ [color=magenta, mark=diamond] table [col sep=comma, x=p, y=FER] {Figures/Short_MultiSeq/bcjr_CC_polar_a1_k32_n64_q4_l32_c8_ker-31_M5.csv};
\addlegendentry{ $M=5$, sep.};

\addplot+ [color=orange, mark=star] table [col sep=comma, x=p, y=FER] {Figures/Short_MultiSeq/bcjr_CC_polar_a1_k32_n64_q4_l32_c8_ker-31_M10.csv};
\addlegendentry{ $M=10$, sep.};

\addplot+ [color=orange, mark=star,dashed, mark options={solid}] table [col sep=comma, x=p, y=FER] {Figures/Short_MultiSeq/BCJR-MSA_CC_polar_a1_k32_n64_q4_l32_c8_ker-31_M10.csv};

\addplot+ [color=teal, mark=o, dashed, mark options={solid}] table [col sep=comma, x=p, y=FER] {Figures/Short_MultiSeq/BCJR-MSA_CC_polar_a1_k32_n64_q4_l32_c8_ker-31_M3.csv};

\addplot+ [color=magenta, mark=diamond, mark options={solid}, dashed] table [col sep=comma, x=p, y=FER] {Figures/Short_MultiSeq/BCJR-MSA_CC_polar_a1_k32_n64_q4_l32_c8_ker-31_M5.csv};

\end{semilogyaxis}%
\end{tikzpicture}%
    \caption{FER performance vs. $p_\I=p_\D$  on short sequences of $\len = 128$ DNA symbols, with $p_\S = 0$ and with no iterations between the inner and outer decoders. The outer code is a polar code with parameters $\leno = 64$, $\rateo=\nicefrac{1}{2}$, $\outq=4$, $\nicefrac{\alpha}{\beta} = 3$, list size $32$, and $\crclen = 8$, while we use the CC-$1$ inner coding scheme. Comparison on different multiple sequence decoding approaches. Solid lines represent separate decoding, dashed are for MSA decoding, and dash dotted are for joint decoding.}
    \label{fig:short-multi-seq}
\end{figure}
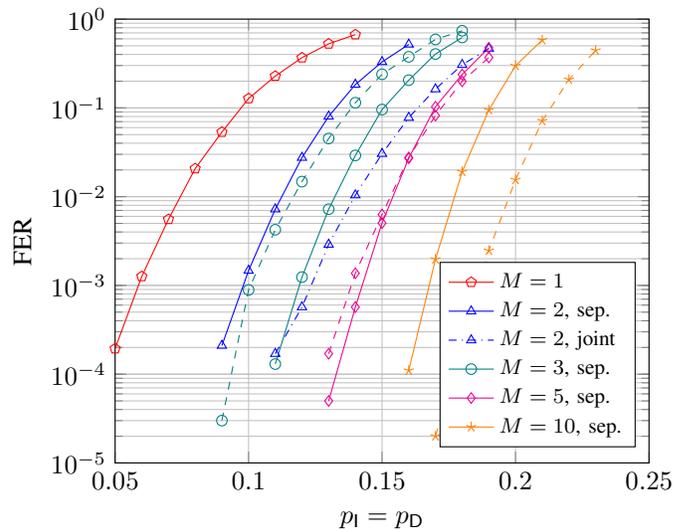

MSA tools are often used in the biological sector for reconstruction of an original sequence given multiple corrupted sequences. Here, we consider the \emph{T-Coffee} MSA method presented in \cite{notredame2000tcoffee} using the open-source library \emph{SeqAn} \cite{Reinert2017seqan}. Given the $\numS$ sequences $\y_1,\ldots,\y_\numS$ from the channel output, the MSA algorithm computes a consensus sequence $\widetilde{\y} = (\widetilde{y}_1,\dots,\widetilde{y}_{\widetilde{\len} })$, where $\widetilde{y}_i \in \Sigma_q$ and $\widetilde{\len}$ is not necessarily equal to $\len$. We use a simple Levenshtein distance orientated scoring scheme to compute the alignment, where a match, a mismatch, and a gap have scores  $0$, $-2$, and $-1$, respectively. Given the computed alignment, we apply a majority decision to output the consensus sequence $\widetilde{\y}$. Subsequently, the sequence $\widetilde{\y}$ is  given to the inner decoder to compute the symbol APPs $p(w_i|\widetilde{\y})$ for each outer codeword symbol as described in Section~\ref{sec:innerdec-one}. We can directly determine the total number of computed operations by the sum of that of decoding  a single received sequence using the inner code and that of the MSA algorithm itself. Therefore, combining the complexity of the T-Coffee algorithm according to \cite{notredame2000tcoffee} and that of decoding a single received sequence using the inner code, the complexity of the MSA method is
\begin{align*}
    C_{\mathrm{MSA}} &= C_{\mathrm{single}} + \mathcal{O}(\numS ^2 \len ^2) + \mathcal{O}(\numS ^3 \len). 
\end{align*}

Fig.~\ref{fig:short-multi-seq} shows the FER of a $[64,32]_{2^2}$ outer polar code concatenated with the CC-$1$ inner coding scheme for a short block length of $\len = 128$ DNA symbols and multiple received sequences. Specifically, we compare the separate and joint decoding approaches discussed in Section~\ref{sec:innerdec-multiple} and the aforementioned MSA method. First observe that due to complexity reasons joint decoding for $\numS > 2$ becomes infeasible. Moreover, the MSA method is impractical for the case of $\numS = 2$ due to the applied majority decision. For the case of $\numS =2$, we see a significant gain of  joint decoding compared to  separate decoding. This is no surprise since joint decoding exploits the full knowledge of the two received sequences concurrently, however, at the expense of a large trellis increasing exponentially in $\numS$. On the other hand, the separate decoding approach ignores during the inner decoding the fact that the received sequences stem from the same original transmitted sequence. Comparing the MSA method and separate decoding for the case of $M > 2$, we see that the performance of separate decoding is better than that of the MSA method for $M=3$, similar for $M=5$, and worse for $M=10$. This can be explained similarly as before, since the MSA method exploits coherences between the received sequences, albeit without any structural assumption, e.g., knowledge of the codebooks. Nevertheless, the gain of the MSA method comes at the price of a higher complexity, where the dominating factor is $\mathcal{O}(\numS ^2 \len ^2) + \mathcal{O}(\numS ^3 \len)$ compared to separate decoding with $\mathcal{O}(\numS \len ^{\nicefrac{3}{2}})$ complexity, as the total number of drift states $\Delta$ is of order $\mathcal{O}(\sqrt{N})$.

For a better grasp on the difference in complexity between joint and separate decoding, we now provide some numerical examples of the total number of candidates for the HMM state variable $\sigma_i$, denoted by $\sigma_{\mathrm{total}}$, for the case of an inner  convolutional code. For a sequence of length $\len = 960$ and $p_\I = p_\D = 0.15$, $\sigma_{\mathrm{total}} = 2^{\nu} \times 131$  for separate decoding, while $\sigma_{\mathrm{total}} = 2^{\nu} \times 131^M$ for joint decoding. For a sequence of length $\len = 128$, $\sigma_{\mathrm{total}} = 2^{\nu} \times 48$ for separate decoding, while $\sigma_{\mathrm{total}} = 2^{\nu} \times 48^M$ for joint decoding.

\subsection{Frame Error Rate Results With  Substitution Errors} \label{sec:FER:substitutions}

\begin{figure}[t]
    \centering
    \begin{tikzpicture}[scale=1.0]
\begin{semilogyaxis}[
width = 0.96\columnwidth,
xmin=0.02,   xmax=0.18,
ymin=1e-5,	ymax=1,
xticklabel style = {/pgf/number format/fixed, /pgf/number format/precision=6},
ymode=log,
grid = both,
grid style = {line width=.1pt},
legend cell align={left},
legend style={font=\footnotesize,at={(axis cs: 0.178,1.5E-5)},anchor=south east},
xlabel = {$p_\I=p_\D$},
ylabel = {FER},
cycle list name=color list
]

\addplot+ [color=blue, mark=square] table [col sep=space, x=pins, y=FER] {Figures/FER_subsErrors/BCJR_CC_optLDPC_q16_n240_k120_ps0.000000_M1.txt};
\addlegendentry{$p_\S = 0$};

\addplot+ [color=red, mark=pentagon] table [col sep=space, x=pins, y=FER] {Figures/FER_subsErrors/BCJR_CC_optLDPC_q16_n240_k120_ps0.050000_M1.txt};
\addlegendentry{$p_\S = 0.05$};

\addplot+ [color=orange, mark=triangle] table [col sep=space, x=pins, y=FER] {Figures/FER_subsErrors/BCJR_CC_optLDPC_q16_n240_k120_ps0.100000_M1.txt};
\addlegendentry{$p_\S = 0.1$};

\addplot+ [color=blue, mark=square, dashed] table [col sep=space, x=pins, y=FER] {Figures/FER_subsErrors/BCJR_CC_optLDPC_q16_n240_k120_ps0.000000_M2.txt};

\addplot+ [color=red, mark=pentagon, dashed] table [col sep=space, x=pins, y=FER] {Figures/FER_subsErrors/BCJR_CC_optLDPC_q16_n240_k120_ps0.050000_M2.txt};

\addplot+ [color=orange, mark=triangle, dashed] table [col sep=space, x=pins, y=FER] {Figures/FER_subsErrors/BCJR_CC_optLDPC_q16_n240_k120_ps0.100000_M2.txt};

\end{semilogyaxis}%
\end{tikzpicture}%
    \caption{FER performance vs. $p_\I=p_\D$ for the CC-$2$ inner coding scheme concatenated with an optimized $[240,120]_{2^4}$  outer LDPC  code with  block length $\len = 960$ DNA symbols for several substitution error probabilities and with   no iterations between the inner and outer decoders. Solid lines correspond to $\numS = 1$ and  dashed lines to $\numS = 2$ with separate decoding.}
    \label{fig:FER_subsErrors}
\end{figure}
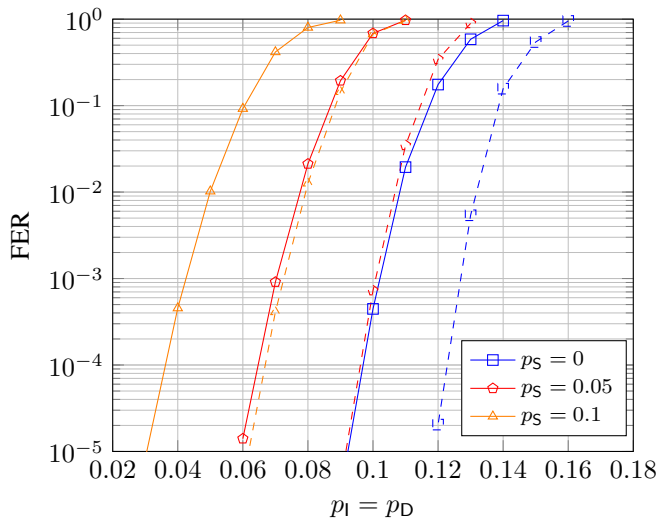

To show the robustness of our coding schemes and decoding algorithms to substitution errors, we show in Fig.~\ref{fig:FER_subsErrors}  FER  results with $p_\S>0$. We use the CC-$2$ inner coding scheme concatenated with an optimized (for single sequence transmission and with   no iterations between the inner and outer decoders) nonbinary protograph LDPC code using the techniques mentioned earlier. The optimal base matrix  for the case of $p_\S = 0$ is $\bm B = \left( \begin{smallmatrix} 1 & 2 & 1 & 1 \\ 1 & 1 & 2 & 1\end{smallmatrix}\right)$ as mentioned earlier, while $\bm B = \left( \begin{smallmatrix} 1 & 2 & 1 & 1 \\ 1 & 1 & 1 & 1\end{smallmatrix}\right)$ gives the best performance  for $p_\S = 0.05$ and $0.1$. The FER curves in Fig.~\ref{fig:FER_subsErrors} are in agreement with the BCJR-once rates in Fig.~\ref{fig:BCJR-Once-Multiple-Ps};  the loss in performance with increasing $p_{\mathsf{S}}$ is approximately the same for $M=1$ and $M=2$ with separate decoding.

%
\section{Conclusion}
In this paper, we proposed concatenated coding schemes for the problem of transmitting one DNA sequence over multiple parallel IDS channels. First, we proposed two novel approaches, or decoding algorithms, for multiple sequence transmission. The first algorithm, being a benchmark for the second one, is an optimal symbolwise MAP decoder but it suffers from high complexity, while the second algorithm is a sub-optimal but more practical decoder with much reduced complexity. We showed with these algorithms that we can achieve significant gains as compared to the single sequence transmission case. Second, we proposed optimization techniques for both the inner and outer codes tailored to the IDS channel. We designed an inner TVC and outer nonbinary LDPC and polar codes that improve the overall performance of the scheme. In addition, we studied the asymptotic performance of different inner codes through AIRs and showed that the FER results are in accordance with them. Lastly, we would like to point out that the code rate for our concatenated coding scheme was chosen for convenience, while for a real-life scheme it should be selected based on the target IDS channel error rates which depend on the sequencing technology. Although it may be considered to be a low rate, the scheme can be straightforwardly adapted to higher rates. In this work, we showed that for an IDS channel with a very high error rate for insertions and deletions, our coding scheme with rate $\nicefrac{1}{2}$ performs very well.
%


%
\appendices
\section{Symbolwise APPs for Memoryless Channels} \label{app:app:memoryless}
We show that for independent channel input symbols, i.e.,  $p(\x) = \prod_{i=1}^{N}p(x_i)$ and a memoryless channel that produces $\numS$ output sequences it holds that
\begin{align*}
    p(x_i|\y) \propto {p(x_i)^{1-\numS}}{\prod_{j=1}^{\numS} p(x_i|\y_j)},
\end{align*}
where the proportionality is with respect to a constant that does not depend on $x_i$. We start with expanding the APP to
\begin{align*}
    &~p(x_i|\y) \overset{(a)}{=} \sum_{\x:x_i} p(\x|\y) = \sum_{\x:x_i} \frac{p(\y|\x) p(\x)}{p(\y)} \\
    &= \sum_{\x:x_i} \frac{p(\x)}{p(\y)} \prod_{j=1}^{\numS} p(\y_j|\x) \\
    &\overset{(b)}{=}  \sum_{\x:x_i} \frac{1}{p(\y)} \prod_{k=1}^{N} p(x_k) \prod_{j=1}^{\numS}  p(y_{j,k}|x_k) \\
    &\overset{(c)}{=} p(x_i)\prod_{j=1}^{\numS}p(y_{j,i}|x_i) \sum_{\x:x_i} \frac{1}{p(\y)} \prod_{k\neq i} p(x_k) \prod_{l=1}^{\numS}  p(y_{l,k}|x_k) \\
    &\propto p(x_i)\prod_{j=1}^{\numS}p(y_{j,i}|x_i),
\end{align*}
where in $(a)$ we used the notation of the sum over $\x:x_i$, which ranges over all vectors $\x$ whose $i$-th symbol is equal to $x_i$. Equality $(b)$ is due to the fact that the channel is memoryless. Finally, in $(c)$ we factored out the terms corresponding to $x_i$, which is possible as $x_i$ is constant within the sum. We can do the analogous steps for $p(x_i|\y_j)$ to deduce that
$$ p(x_i|\y_j) \propto p(x_i) p(y_{j,i}|x_i).$$
Combining these two equivalences, we arrive at the desired proportionality. Notice that this relation translates to the APPs $p(w_i|\y)$ when no inner code is used, i.e., when $\w=\x$.

\ifCLASSOPTIONcaptionsoff
  \newpage
\fi
\balance


\end{document}